\documentclass[%
 reprint,
 amsmath,amssymb,
aps,
pra,
superscriptaddress,
longbibliography,
]{revtex4-2}
\usepackage{multirow}
\usepackage{graphicx}
\usepackage{dcolumn}
\usepackage{bm}
\usepackage[version=4]{mhchem}
\usepackage{braket}
\usepackage{xcolor}
\usepackage{soul}

\usepackage[T1]{fontenc} 

\begin{document}

\preprint{APS/123-QED}
\title{Exploring the feasibility of  probabilistic and deterministic quantum gates between T centers in silicon}

\author{Shahrzad Taherizadegan} 
\affiliation{Department of Physics \& Astronomy, Institute for Quantum Science and Technology, University of Calgary, 2500 University Drive NW, Calgary, Alberta T2N 1N4, Canada}
\author{Faezeh Kimiaee Asadi}
\affiliation{Department of Physics \& Astronomy, Institute for Quantum Science and Technology, University of Calgary, 2500 University Drive NW, Calgary, Alberta T2N 1N4, Canada}
\author{Jia-Wei Ji}
\affiliation{Department of Physics \& Astronomy, Institute for Quantum Science and Technology, University of Calgary, 2500 University Drive NW, Calgary, Alberta T2N 1N4, Canada}
\author{Daniel Higginbottom}
\affiliation{Department of Physics, Simon Fraser University, Burnaby, BC V5A 1S6, Canada}
\author{Christoph Simon}
\affiliation{Department of Physics \& Astronomy, Institute for Quantum Science and Technology, University of Calgary, 2500 University Drive NW, Calgary, Alberta T2N 1N4, Canada}


\begin{abstract}
T center defects in silicon provide an attractive platform for quantum technologies due to their unique spin properties and compatibility with mature silicon technologies. We investigate several gate protocols between single T centers, including two probabilistic photon interference-based schemes, a near-deterministic photon scattering gate, and a deterministic magnetic dipole-based scheme. In particular, we study a photon interference-based scheme with feedback which can achieve success probabilities above $50\%$, and use the photon-count decomposition method to perform the first analytical calculations of its entanglement fidelity and efficiency while accounting for imperfections. We also calculate the fidelity and efficiency of the other schemes. Finally, we compare the performance of all the schemes, considering current and near-future experimental capabilities. In particular, we find that the photon interference-based scheme with feedback has the potential to achieve competitive efficiency and fidelity, making it interesting to explore experimentally. 

\end{abstract}
\maketitle


\section{Introduction}
\begin{figure}[h!]
\centering
\includegraphics[scale=0.2]{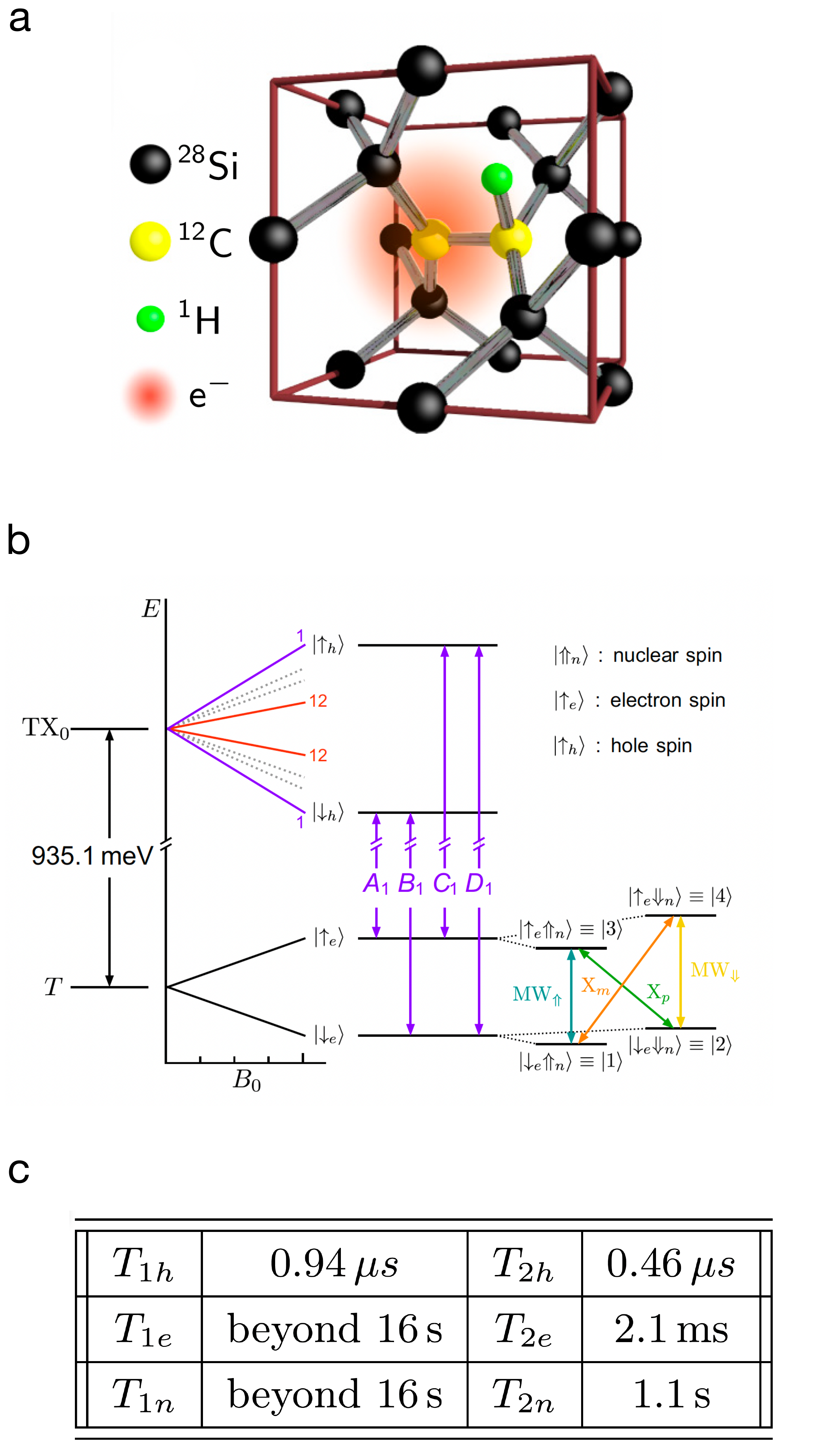}
\caption{\textbf{(a)} shows the atomic structure of the T center in silicon based on the proposal of Ref. \cite{PhysRevLett.77.4812}. Figure from Ref. \cite{bergeron2020silicon}. \textbf{(b)} Energy level structure of T center and the optical transitions $A_1, B_1, C_1, D_1$ of the $\mathrm{{TX}_0}$ bound exciton excited state. The anisotropy of the unpaired hole spin in the bound exciton state results in $12$ independently addressable orientational subsets. Figure from \cite{higginbottom2023memory}. \textbf{(c)} Optical excited state and ground state properties of T centers in bulk silicon \cite{bergeron2020silicon, higginbottom2023memory}. $T_1$ and $T_2$ represent the lifetimes and coherence times, respectively, while $h$, $e$, and $n$ denote the hole spins, electron spins, and nuclear spins.}
\label{Tcenter}
\end{figure}

Quantum gates and entanglement generation schemes are foundational to realize future quantum technologies. They are essential for processing quantum information in quantum computers \cite{nielsen2010quantum, deutsch1989quantum, ladd2010quantum} and for the generation, storage, and swapping of entanglement for long-distance quantum communication through quantum repeaters \cite{sangouard2011quantum, duan2001long}, paving the way for building the future global quantum internet \cite{simon2017towards, kimble2008quantum, wehner2018quantum}.
\\\indent Solid-state systems have attracted considerable attention as a platform for future quantum networks elements, including quantum memories \cite{ruf2021quantum, pompili2021realization, knaut2023entanglement, lei2023quantum, taherizadegan2024towards}. Quantum gates and entanglement generation schemes have been proposed and implemented in various solid-state systems, such as rare-earth ions doped in crystals \cite{grimm2021universal, ohlsson2002quantum, asadi2020protocols}, nitrogen-vacancy (NV) centers \cite{bernien2013heralded}, and silicon-vacancy (SiV) centers in diamond \cite{evans2018photon, day2022coherent}.
\\\indent The family of silicon defects have been investigated for decades starting even before the development of quantum technology \cite{safonov1996interstitial, safonov1999photoluminescence, minaev1981thermally, irion1985defect, safonov1993hydrogen, leary1998interaction}. These defects, which exhibit sharp spectral features, are produced by irradiation or heat treatment, and are therefore often referred to as radiation damage centers. The T center in silicon has recently drawn significant attention in the context of quantum technology due to its unique features \cite{simmons2024scalable}. Silicon is an attractive host due to its compatibility with the current established electronics and photonics platforms. The T center in silicon offers long lived electron and nuclear spins, and, importantly, spin-selective bound exciton excited state optical transitions at $1326\, \mathrm{nm}$ wavelength in the telecommunication O-band \cite{bergeron2020silicon} making it an ideal matter-photon interface. The T center in isotopically purified $\mathrm{^{28}Si}$ offers enhanced electron and nuclear spin lifetimes and reduced inhomogeneous broadening of optical transitions. 
\\\indent T centers have also been created and studied in commercial materials other than bulk silicon such as the $220\, \mathrm{nm}$ device layer of silicon-on-insulator wafers \cite{macquarrie2021generating,higginbottom2022optical} and integrated silicon photonic waveguides \cite{deabreu2023waveguide, komza2025multiplexed}.
\\\indent Besides, some of us have explored the potential of T center ensembles for implementing various quantum memory and transduction schemes by characterizing the T center spin ensemble, measuring optical depth, and discussing achievable efficiency \cite{higginbottom2023memory}.   
\\\indent Figure \ref{Tcenter}\,\textbf{a} illustrates the atomic structure of the T center in silicon, as proposed by \cite{PhysRevLett.77.4812} and later confirmed through observations in \cite{bergeron2020silicon} and a first-principles study in \cite{dhaliah2022first}. Figure \ref{Tcenter}\,\textbf{b} depicts the level structure and relevant optical transitions of the T center. The characteristics of the ensemble of T centers in bulk silicon are shown in figure \ref{Tcenter}\,\textbf{c}.
\\\indent So far, Ref. \cite{afzal2024distributed} has shown the first demonstration of entanglement generation between the electron spins of two remote individual T centers using the photon interference-based scheme proposed by Barrett and Kok \cite{barrett2005efficient}. A CNOT gate in a single T center, with the control and target being the nuclear and electron spins $\mathrm{(C_n NOT_e)}$ was also demonstrated by Rabi oscillations driving the $\mathrm{MW_{\Downarrow}}$ transition \cite{afzal2024distributed}. Assuming enhancements in the fabrication of integrated T centers and that the bulk T center properties represent the ultimate performance of integrated devices, Ref. \cite{afzal2024distributed} also evaluated the potential achievable efficiency and the maximum fidelity of performing the interference-based Barrett-Kok entanglement distribution protocol for future T center systems.
\\\indent The gate scheme discussed in Ref. \cite{afzal2024distributed} is probabilistic, with a maximum theoretical success probability of $50\%$. Considering the limitations of current experimental capabilities, the practical success rate is lower, imposing even greater overheads for various applications. This raises interest in exploring alternative schemes that could either be deterministic or achieve higher success probabilities given current and near-future experimental potential.
In particular, implementing feedback during the measurement based on photon counting following the photon interference-based scheme as proposed in \cite{martin2019single} promises a higher success probability compared to the scheme discussed in Refs. \cite{afzal2024distributed, barrett2005efficient} and unit efficiency in the ideal case.
\\\indent Here, we analyze this scheme and apply the photon-count decomposition method described in Ref. \cite{wein2020analyzing} to derive analytical expressions for its fidelity and efficiency, accounting for various imperfections, including optical decoherence, spin dephasing, photon loss, and collection and detection inefficiencies. We use our results to evaluate the performance of this scheme and explore the feasibility of implementing other schemes for T centers. We quantitatively analyze these schemes based on previous works \cite{wein2020analyzing, asadi2020cavity, karimi2024comparing}, which were originally developed for different systems. Finally, we compare their performance in light of current experimental advancements. Our findings suggest that the photon interference-based scheme with feedback offers a competitive combination of efficiency and fidelity, making it worth investigating experimentally.
\\\indent We discuss implementing probabilistic protocols based on photon interference, both with and without feedback in section \ref{probabilistic photon-mediated gates}. Additionally, we examine a near-deterministic photon scattering-based gate in section \ref{SB} and a deterministic magnetic dipole-based gate in section \ref{MDG}. For each scheme, we estimate key characteristics, such as fidelity, efficiency, and gate time. In section \ref{comparison}, we compare the performance of the schemes. Finally, the conclusion and outlook are given in section \ref{conclusion}. 



\section{probabilistic photon interference-based gates}
\label{probabilistic photon-mediated gates}
First, we discuss the photon interference-based (IB) scheme in subsection \ref{IB scheme}. Next, in subsection \ref{IBF}, we quantify the performance of the IB scheme with feedback (IBF). It is worth noting that these entanglement generation protocols can be transformed into a controlled-Z (CZ) gate by changing the measurement basis to a mutually unbiased basis \cite{lim2005repeat}.

\subsection{INTERFERENCE-BASED SCHEME}
\label{IB scheme}
The interference-based (IB) entanglement generation scheme, introduced by Barrett and Kok, is widely used in quantum networks and distributed quantum computing \cite{barrett2005efficient}. It has been demonstrated experimentally for entanglement generation in several systems e.g., two trapped ytterbium ions in vacuum chambers separated by $1\, \mathrm{m}$ \cite{moehring2007entanglement}, ytterbium ions in nanophotonic crystal cavities \cite{ruskuc2024scalable}, and nitrogen–vacancy (NV) centers in diamond with a separation of $3\,\mathrm{m}$ \cite{bernien2013heralded}. Recently, this scheme has been demonstrated experimentally in T centers separated by $6 \,\mathrm{m}$ \cite{afzal2024distributed}. 
\\\indent Let us consider two three-level atoms with long-lived ground states $\ket{\uparrow}$, and $\ket{\downarrow}$ (the electron spins), and an excited state $\ket{e}$ (the hole spin) which decays to $\ket{\downarrow}$. This scheme is implemented by preparing the atoms in the superposition of their ground states, $(\ket{\uparrow} + \ket{\downarrow}) / \sqrt{2}$. Then, an optical $\pi$ pulse is applied to each qubit, coherently driving the $\ket{\downarrow}\leftrightarrow\ket{e}$ transition. The resulting emissions from both systems are interfered on a beam splitter, and the system is left to wait until the detection time $T_d$ to detect exactly one photon. Assuming the photons emitted by the two atoms are indistinguishable, the two atoms will be projected onto one of the maximally entangled Bell states. Due to photon loss, it is possible in reality to detect a single photon even when both atoms are in the state $\ket{\downarrow \downarrow}$. To remove such a possibility both qubits' spins are flipped and the process is repeated. The state $\ket{\downarrow \downarrow}$ is flipped to $\ket{\uparrow \uparrow}$ and will emit no photons. Detection of another photon in the second round heralds the entanglement generation \cite{barrett2005efficient, bernien2013heralded}.
\\\indent Thus, the entanglement time for the IB scheme is approximately $T_{\mathit{IB}} = 2\,T_d+\delta t$ where $\delta t$ is the time to perform the spin flip. Since the final maximally entangled state can be one of two Bell states out of the four Bell components of the initial state, the maximum theoretical efficiency of the scheme is $\eta_{\mathit{IB}} = 0.5$. The efficiency of Bell state measurement can in principle be improved beyond $0.5$ by employing methods such as the use of ancillary photons \cite{grice2011arbitrarily, ewert20143}. The performance of the scheme employing ancillary photons has been quantified and compared to conventional Bell state measurements \cite{wein2016efficiency}, and its feasibility has been experimentally demonstrated \cite{bayerbach2023bell}. The efficiency can approach unity by utilizing increasingly complex ancillary states. However, this comes at the cost of elevated apparatus complexity, a higher likelihood of component errors, and the need for multi-photon detection, all of which restrict their practical application.
\\\indent The efficiency and fidelity of the IB entanglement generation scheme can be calculated employing the photon-count decomposition method \cite{wein2020analyzing} where the density operator solution of the master equation is decomposed into a set of conditional propagation superoperators dependent on the cumulative detector photon count. 
\\\indent While this gate scheme does not inherently require a cavity for operation, its fidelity rapidly decreases with increasing detection time in the absence of a cavity, and the efficiency becomes too low. We show that incorporating a cavity significantly enhances both the efficiency and fidelity. Each system can be placed in a separate optical cavity, with only the transition $\ket{\downarrow} \leftrightarrow \ket{e}$ coupled to the cavity mode.
\\\indent When the gate time approaches or exceeds the spin coherence time, spin decoherence begins to affect the system. Since we use the electron spin in single T centers, the dominant spin decoherence comes from spin dephasing, the rate of which is four orders of magnitude larger than that of spin decay, as shown in figure \ref{Tcenter}\,\textbf{c}. In this case, we can consider only the effect of spin dephasing to quantify the efficiency and fidelity. The efficiency of the IB scheme employing the cavity is given by 

\begin{equation}
\mathrm{\eta_{IB}} = \frac{\eta'^2}{2} (1-e^{-T_d\gamma'})^2
\label{etaIBcavity}
\end{equation}
where $\gamma'$ is the Purcell enhanced optical decay rate, the term $e^{-T_d\gamma'}$ describes the probability of the T center remaining in its excited state after a time $T_d$, and $\eta'$ is 
\begin{equation}
\eta' = \eta_d\, \eta_c \, \frac{F_p\gamma_{zpl}}{\gamma'}.  
\label{etaprime}
\end{equation}
In this expression, $\eta_d$ and $\eta_c$ represent the detection and collection efficiencies, respectively. The zero-phonon line (ZPL) emission rate is given by $\gamma_{\mathit{zpl}} = \gamma_r \eta_{\mathit{zpl}}$, where the ZPL efficiency ($\eta_{\mathit{zpl}}$)—also known as the Debye-Waller factor—is the fraction of total emission occurring in the ZPL. For T centers, this value is $\eta_{\mathit{zpl}} = 0.23$ \cite{bergeron2020silicon}. The radiative decay rate $\gamma_r$ can be expressed as $\gamma_r = \eta_r \gamma$, where $\gamma = 1/T_{1h}$ is the optical decay rate between $\ket{e}$ and $\ket{\uparrow}$, and $\eta_r$ is the radiative efficiency. While the exact value of $\eta_r$ is not yet precisely known, \cite{johnston2024cavity} established a lower bound of $\eta_r \geq 0.23$ for an individual T center, and, for all gate performance evaluations, we assume $\eta_r = 0.23$. The Purcell factor, $F_p$, describes the ZPL cavity enhancement. Due to the significant presence of phonon sidebands in T centers, the modified decay rate in a cavity with Purcell enhancement is given by $\gamma' = F_p \gamma_{\mathit{zpl}} + \gamma$. The term $\frac{F_p\gamma_{\mathit{zpl}}}{\gamma'}$ in the equation \eqref{etaprime} is the photon emission efficiency from the cavity mode $(\eta_{em})$ in the bad cavity regime with $\gamma \ll \kappa$ \cite{grange2015cavity,asadi2020protocols}. Recently, a value of $\gamma'= 2\pi\times2.5 \, \mathrm{MHz}$ was achieved experimentally by incorporating individual T centers into an optical cavity on a photonic chip \cite{afzal2024distributed}, resulting in an emission efficiency of $\eta_{em} = 0.93$ for a cavity-coupled single T center. 
\\\indent The fidelity of the IB entanglement generation scheme within the cavity is calculated as

\begin{equation}
F_{\text{IB}}=\frac{1}{2}\left(1+\frac{(\Tilde{C}_{\text{IB}}(T_d))^2}{(1-e^{-\gamma'T_d})^2}\right), 
\label{FIBcavity}
\end{equation}
with
\begin{equation}
\Tilde{C}_{\text{IB}}(T_d)=\frac{\gamma'}{\Gamma'+\gamma^*_s}e^{-2\gamma^*_s\delta t}(1-e^{-(\Gamma'+\gamma^*_s)T_d}).   
\label{ctildecavity}
\end{equation}
$\Gamma' = \gamma' +  2\gamma^*$ represents the FWHM of the Purcell-enhanced emission line, where $\gamma^* = 1/T_{2h} - \gamma/2$ is the optical pure dephasing rate, and $\gamma^*_s = 1/T_{2e} - 1/2T_{1e}$ is the spin dephasing rate. $\delta t$ is the time for the spin flip.
\\\indent Since we consider a local gate, we assume no delay in detection in both rounds, meaning the time window between emission and detection is set to zero (see Section \ref{IBF} for more details).   

\begin{figure*}[t]
\centering
\includegraphics[scale=0.40]{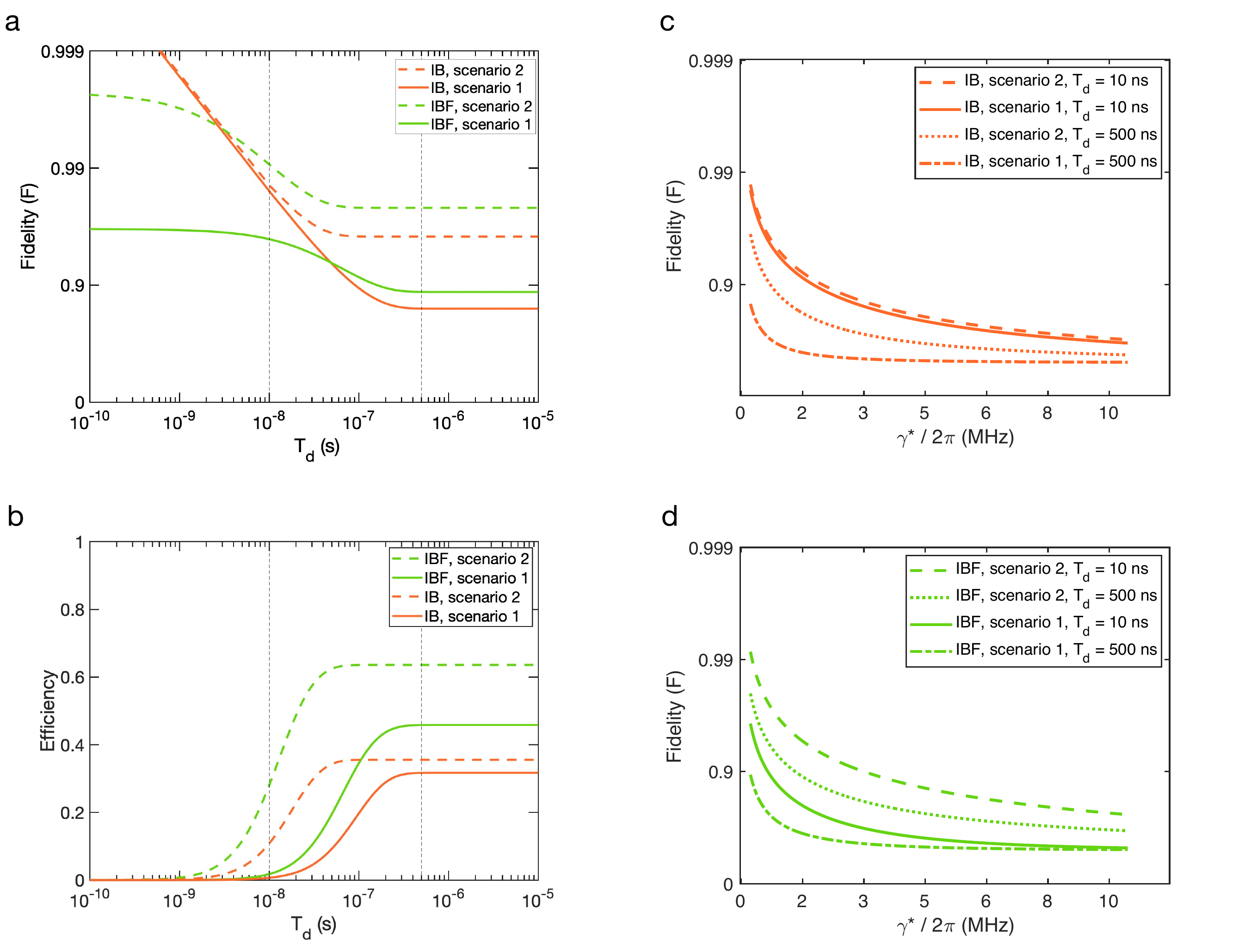}
\caption{The figure compares the fidelity and efficiency of the IB and IBF schemes across different scenarios as a function of detection time $T_d$ (\textbf{a} and \textbf{b}), and shows the fidelity of the schemes as a function of the optical dephasing rate $\gamma^*$ (\textbf{c} and \textbf{d}). Each color corresponds to a certain scheme. Scenario 1 uses the currently demonstrated experimental values for the Purcell-enhanced optical decay rate $\gamma' = 2\pi\times2.5 \, \mathrm{MHz}$, Purcell factor $F_p = 256.5$, and delay time $\delta t = 20.9 \, \mathrm{ns}$, whereas scenario 2 considers the more optimistic values of $\gamma' = 2\pi\times12.7 \, \mathrm{MHz}$, $F_p = 1402.4$, and $\delta t = 1.4 \, \mathrm{ns}$. Optical dephasing is set to the bulk value in both cases \cite{higginbottom2023memory}. In \textbf{(c)} and \textbf{(d)}, the $\gamma^*$ extends from the optimal value achieved in bulk \cite{higginbottom2023memory} to the value measured in device \cite{afzal2024distributed}. For both the IB and IBF schemes, the results are shown at two detection times, indicated by the vertical dashed lines in \textbf{(a)} and \textbf{(b)} (see the discussion in section \ref{probabilistic photon-mediated gates}).}
\label{FetavsTd}
\end{figure*}

We evaluate the performance of the IB scheme in two main scenarios. Scenario 1 uses the currently demonstrated experimental values for the Purcell-enhanced optical decay rate $\gamma'= 2\pi\times2.5 \, \mathrm{MHz}$ and Purcell factor $F_p = 256.5$ \cite{johnston2024cavity, afzal2024distributed, islam2023cavity, komza2025multiplexed} while scenario 2 considers the more optimistic values of $\gamma'=2\pi\times12.7 \, \mathrm{MHz}$ \cite{higginbottom2023integrated} and $F_p=1402.4$ that are thought to be attainable experimentally in the near future. Note the $\gamma' = 2\pi\times12.7 \, \mathrm{MHz}$ is based on the estimation of Purcell enhanced emission rate in T center device \cite{higginbottom2023integrated}. Given that a collection efficiency of $\eta_c = 0.7$ has been achieved for a T center in a silicon nanobeam waveguide \cite{lee2023high}, we assume $\eta_c = 0.9$ for all gate performance evaluations within the cavity. The detection efficiency is assumed to be $\eta_d = 0.95$, which is achievable with current technology, as similar or higher values have been demonstrated for nearby wavelengths using superconducting nanowire single-photon detectors (SNSPDs) \cite{chang2021detecting, reddy2020superconducting, hu2020detecting}. The fidelity and efficiency of the IB entanglement generation scheme employing a cavity for both scenarios are shown in figure \ref{FetavsTd} (\textbf{a} and \textbf{b}). As anticipated, achieving the optimistic values leads to notable enhancements in both fidelity and efficiency. These results will be discussed in more detail at the end of the next subsection.
\\\indent Figure \ref{FetavsTd} \textbf{c} shows the fidelity of the IB scheme as a function of the optical dephasing rate, ranging from values characteristic of device-integrated T centers \cite{deabreu2023waveguide, afzal2024distributed} to those achieved in bulk $\mathrm{^{28}Si}$ \cite{higginbottom2023memory}. Both scenarios are included, along with two representative detection times chosen to highlight the trade-off between efficiency and fidelity (see the next subsection for further discussion).

\subsection{INTERFERENCE-BASED SCHEME WITH FEEDBACK}
\label{IBF}
Here, we consider the interference-based scheme with feedback (IBF), which was originally proposed in \cite{martin2019single} as a modified version of the IB scheme. We derive the general analytical efficiency and fidelity equations for this scheme and apply them to T centers to analyze the performance of the scheme. While the scheme is not deterministic in practice, it can achieve an efficiency greater than $0.5$, depending on the system's capabilities. 
\\\indent Similar to the IB scheme, two atoms, each with three levels — one excited state $\ket{e}$ and two ground states $\ket{\uparrow}$ and $\ket{\downarrow}$ — are required. Both systems are initialized in their excited state $\ket{e}$ $(\psi_0 = \ket{ee})$. The excited state $\ket{e}$ decays into the state $\ket{\downarrow}$. The spontaneous emissions from both systems are merged at a beam splitter, erasing any information about which photon originated from which atom. After detecting one photon at $\tau_1$, the state of the system would be $\psi(\tau_1) = (\ket{e\downarrow} \pm \ket{\downarrow e})/\sqrt{2}$. Then applying a $\pi$ pulse between the ground states $\ket{\downarrow}$ and $\ket{\uparrow}$, which takes a time $\delta t$ and is called feedback, results in the state $\psi(\tau_{pulse}) = (\ket{e\uparrow} \pm \ket{\uparrow e})/\sqrt{2}$ for the system. Detection of the second photon at $\tau_2$ projects the state of the system into the maximally entangled Bell state $\psi(\tau_2) = (\ket{\downarrow\uparrow} \pm \ket{\uparrow\downarrow})/\sqrt{2}$ (see figure $1$ in \cite{martin2019single}). Ideally, this scheme can achieve unit efficiency; however, it is constrained by the non-zero time required to perform the feedback and the possibility of the atom decaying before the $\pi$-pulse is applied.
\\\indent As in \ref{IB scheme}, embedding individual T centers in the cavity enhances the efficiency and fidelity of the IB scheme. Therefore, we employ the cavity in the IBF scheme, and the cavity is coupled to the transition from $\ket{e}$ to $\ket{\downarrow}$.
\\\indent Applying the photon-count decomposition method described in \cite{wein2020analyzing}, one can quantify the entanglement generation efficiency and fidelity of the IBF scheme. Similar to the IB scheme, there are four possible photon counts: $\mathbf{n}=\{\mathbf{n}_l,\mathbf{n}_\text{e}\}=\{(1,0),(1,0)\}$, $\{(1,0),(0,1)\}$, $\{(0,1),(1,0)\}$, and $\{(0,1),(0,1)\}$ where $\mathbf{n}_l$ and $\mathbf{n}_e$ stand for the photon count in the early and late detection time window, and each can take two possible outcomes $(1,0),(0,1)$ which correspond to the click in the left detector and the right detector. Thus, the entanglement generation efficiency and fidelity are defined in the following way 
\begin{align}
\eta_{\text{gen}}&=\text{Tr}[\hat{\rho}(t_f)]=\sum_{\mathbf{n}}^{}{\text{Tr}[\hat{\rho}_{\mathbf{n}}(t_f)]}\\
F_{\text{gen}}&= \frac{1}{4}\sum_{\mathbf{n}}^{}{\frac{\bra{\psi_\pm}\hat{\rho}_{\mathbf{n}}(t_f)\ket{\psi_\pm}}{\text{Tr}[\hat{\rho}_{\mathbf{n}}(t_f)]}},
\end{align}

where we use $\ket{\psi_+}$, when $\mathbf{n}=\{(1,0),(1,0)\}, \{(0,1),(0,1)\}$, otherwise we use $\ket{\psi_-}$. The unnormalized conditional state $\hat{\rho}_{\mathbf{n}}(t_f)$ corresponds to the state conditioned on the measurement outcome at the final protocol time $t_f$, which is given by
\begin{align}
&\hat{\rho}_{\mathbf{n_e}}(T_d)=\int_{0}^{T_d}{e^{L_c\delta t_d}\mathcal{S}_{\mathbf{n_e}}e^{L_c t_1}\rho_0 \, dt_1},\\
&\hat{\rho}_m=e^{L_c\delta t}\mathcal{X}\hat{\rho}_{\mathbf{n_e}}(T_d),\\
&\hat{\rho}_{\mathbf{n}}(t_f)=\int_{T_d+\delta t}^{t_f}{e^{L_c\delta t_d}\mathcal{S}_{\mathbf{n_l}}e^{L_c(t_2-T_d-\delta t)}\hat{\rho}_m\, dt_2},    
\end{align}
where $\hat{\rho}_{\mathbf{n_e}}(T_d)$ is the conditional state at detection time $T_d$, $\rho_0$ is the initial state, and $L_c$ is the Liouville superoperator without containing the collapse operators \cite{wein2020analyzing}. $\delta t_d=T_d-t_1=t_f-t_2$ is the time window between the collapse and detection. Since the gate is local, we set this time window to be zero, i.e., $\delta t_d=0$. $\hat{\rho}_m$ is the state after the spin-flip with $\mathcal{X}$ representing the superoperator propagator that executes the spin-flip on both systems during $\delta t$. $t_f = T_{\text{IBF}} = 2T_d + \delta t$ is the gate time for the IBF scheme. $\mathcal{S}_{\mathbf{n_e}}$ and $\mathcal{S}_{\mathbf{n_l}}$ stand for the collapse operators for the early and late time bin photons. 
\\\indent In the presence of spin dephasing, the efficiency of the IBF scheme is computed as follows
\begin{equation}
\eta_{\text{IBF}}=\eta'^2e^{-\gamma'\delta t}(1-e^{-2\gamma'T_d})(1-e^{-\gamma'T_d}).
\label{etaIBF}
\end{equation}
Both the detection time $T_d$ and delay time $\delta t$ affect the efficiency in the IBF scheme as opposed to the IB scheme which is insensitive to the spin-flip time. In the optical limit $T_d\gg 1/\gamma'$, the efficiency becomes

\begin{equation}
\eta_{\text{IBF}}=\eta'^2 e^{-\gamma'\delta t}.    
\end{equation}

The fidelity is given by the following expression
\begin{equation}
F_{\text{IBF}}=\frac{1}{2}\left(1+\frac{\Tilde{C}_{\text{IBF}}(T_d)}{1-e^{-\gamma'T_d}}\right),    
\end{equation}
with
\begin{equation}
\Tilde{C}_{\text{IBF}}(T_d) =\frac{\gamma'}{\Gamma'+\gamma^*_s}e^{-\delta t(2\gamma^*+\gamma^*_s)}(1-e^{-T_d(\Gamma'+\gamma^*_s)}),  
\end{equation}
\\\indent In the absence of spin dephasing, the fidelity of the scheme can be calculated analytically considering the optical frequency difference between the two T centers, and phase errors (see Supplementary Material section \ref{IBF no spin dephasing}). 
\\\indent To evaluate the scheme's performance in analogy with the IB scheme, we examine $\delta t$ as the time required to perform a spin flip between the ground states, determined by the microwave Rabi frequency ($\delta t = \pi / \Omega$). In Scenario 1, we assume a delay time of $\delta t = 20.9\, \mathrm{ns}$, corresponding to an estimated microwave Rabi frequency of $\Omega = 2\pi\times 23.9 \, \mathrm{MHz}$, which is about an order of magnitude larger than the $\Omega = 2\pi\times 2 \, \mathrm{MHz}$ reported in \cite{afzal2024distributed}. In Scenario 2, we consider a more optimistic delay time of $\delta t = 1.4\, \mathrm{ns}$. Note that by applying the Raman transitions between the ground states \cite{jenkins2022ytterbium}, one can drive Rabi frequencies limited by the splitting between the ground states, $\Delta_{\uparrow \downarrow}$, with $\delta t = \pi / \Delta_{\uparrow \downarrow}$. Here, the electron spin states serve as the ground states, with $\Delta_{\uparrow \downarrow} = 2\pi \times 2.25 \, \mathrm{GHz}$, leading to a delay time of $\delta t = \pi / \Delta_{\uparrow \downarrow} = 0.2 \, \mathrm{ns}$. We note that, in practice, electronic delays also contribute; we will return to this point in the discussion and outlook sections.
\\\indent The fidelity and efficiency of the IBF scheme are shown in figure \ref{FetavsTd}\,\textbf{a} (\textbf{b}). We can see that even with the current experimental values for $\eta_d$ and $\eta_c$ and the lower bound of $\eta_r$, it is possible to achieve efficiencies greater than those attainable by the IB scheme in the same scenario. The fidelity and efficiency can be strongly influenced by different combinations of $\delta t$ and $\gamma'$ values, with scenario 2 showing notable improvements in the scheme’s performance.
\\\indent Figure \ref{FetavsTd}\,\textbf{a} shows that for both photon interference-based schemes, in the presence of spin decoherence, the fidelity decreases rapidly until approximately $T_d = 100 \,\mathrm{ns}$. It then remains constant and stabilizes for longer $T_d$. On the other hand, figure \ref{FetavsTd}\,\textbf{b} shows that the efficiency starts at zero and increases rapidly until approximately $500\,\mathrm{ns}$, after which it approaches saturation in both schemes and scenarios. It reaches its maximum value in the so-called optical limit, where $T_d\gg 1/\gamma'$. Notably, the time at which the efficiency approaches its maximum and the fidelity reaches its plateau differs depending on the specific scenario considered. Thus, there is a trade-off between efficiency and fidelity in photon interference-based schemes. Higher fidelities can be achieved at shorter detection times at the cost of lower efficiencies. Conversely, waiting until approximately $500 \, \mathrm{ns}$ allows for maximum efficiency but results in lower fidelity. The prioritization of efficiency or fidelity depends on the ultimate goal of the application. In section \ref{comparison}, we will discuss the dependence of fidelity and efficiency on the cavity cooperativity at two different detection times.
\\\indent Figure \ref{FetavsTd}\,\textbf{c} and \textbf{d} show the fidelity of the IB and IBF entanglement generation schemes as a function of the optical dephasing rate $\gamma^*$ for both scenarios and at two representative detection times ($T_d = 10\,\mathrm{ns}$ and $T_d = 500\,\mathrm{ns}$) corresponding to the the efficiency–fidelity trade-off. In both schemes, the there is a uniform decrease in the fidelity with increasing $\gamma^*$, as expected due to the detrimental effect of optical dephasing on photon indistinguishability. 
For a fixed $\gamma^*$, shorter detection times yield higher fidelities, in agreement with the above explanations and highlighting the importance of temporal filtering for achieving high fidelities \cite{bowness2025laser, ngan2024performance}. Scenario 2 also outperforms scenario 1 across all schemes and detection times, with higher fidelities throughout the entire range of $\gamma^*$, consistent with the findings in figure \ref{FetavsTd}\,\textbf{a}. For the IB scheme, fidelity drops below $0.9$ for $\gamma^* \gtrsim 2\pi\times2.1\,\mathrm{MHz}$ at short detection times, whereas for the IBF scheme in scenario 2, this limit occurs at a higher $\gamma^*$ value, while in scenario 1, it occurs at a lower $\gamma^*$ value. With continued advancements in host materials and fabrication techniques, the optical dephasing times of device-integrated T centers are expected to improve substantially, approaching those achieved in bulk $\mathrm{^{28}Si}$. 

\section{near-deterministic photon scattering-based scheme}
\label{SB}
In this section, we investigate and analyze quantitatively the cavity-assisted photon scattering-based (SB) scheme for implementation in T centers based on the work done in \cite{asadi2020cavity, Faezehcavityerratum}. 
\\\indent The cavity-assisted photon scattering-based (SB) scheme was originally proposed by Duan and Kimble \cite{duan2004scalable}. Since then various modifications have been proposed based on the main proposal for implementation of local \cite{duan2005robust} and non-local \cite{xiao2004realizing} phase-flip gates. Ref. \cite{asadi2020cavity} has studied the gate in both the bad-cavity regime and the strong-coupling regime, compared to previous works where only the strong-coupling regime is discussed. 
\\\indent The gate is implemented between two qubits located in a single-sided cavity by reflecting a single photon off the qubit-cavity system and detecting it. Detecting the photon after its reflection from the cavity heralds the gate operation. One needs a system with three levels: two ground states ($\ket{\uparrow}$, $\ket{\downarrow}$) and one excited state ($\ket{e}$). The transition $\ket{\uparrow}\leftrightarrow\ket{e}$ is resonant with the cavity mode. We denote the state of the single-photon pulse, which is resonant with the cavity mode, by $\ket{p}$. If both qubits are in the state $\ket{\downarrow}$, the photon enters the cavity, and reflects from inside the cavity with a global phase flip, resulting in a $\pi$-phase shift in the joint state of the qubit-photon system. However, if either or both qubits are in state $\ket{\uparrow}$, the cavity mode is modified due to the atom-cavity coupling. In this case, the photon does not enter the off-resonant cavity and instead reflects off the cavity's out-coupling mirror without any phase shift leading to a controlled-Z (CZ) gate.
\\\indent The gate time is $T_{\mathit{SB}}=8\pi\sqrt{2\ln2}/\sigma_p$, which is twice the FWHM of the
photon duration, where $\sigma_p$ is the spectral standard deviation of the photon with a Gaussian intensity profile. 
Incorporating the detection efficiency $\eta_d$, the efficiency of performing the gate is given by 
\begin{equation}
\eta_{\text{SB}} = \left(1 - \frac{5}{2C} - \frac{(\delta_{\epsilon_{A}}-\delta_{\epsilon_{B}})^2}{2\gamma^2 C}\right) \, \eta_d    
\end{equation}
where $C$ is the cavity cooperativity, and $\delta_{\epsilon_{A}}(\delta_{\epsilon_{B}})$ denotes the detuning between the optical transition of system A (B) (corresponding to the T centers in our case) and the cavity mode. The term in the expression in parentheses represents the probability that the photon is scattered off the emitters inside the cavity \cite{Faezehcavityerratum}, which, in the zero-detuning case, matches the empirical formula for the success probability limited by photon loss due to atomic spontaneous emission, as given in \cite{duan2004scalable, duan2005robust}.
\begin{figure}[h!]
\centering
\includegraphics[scale=0.43]{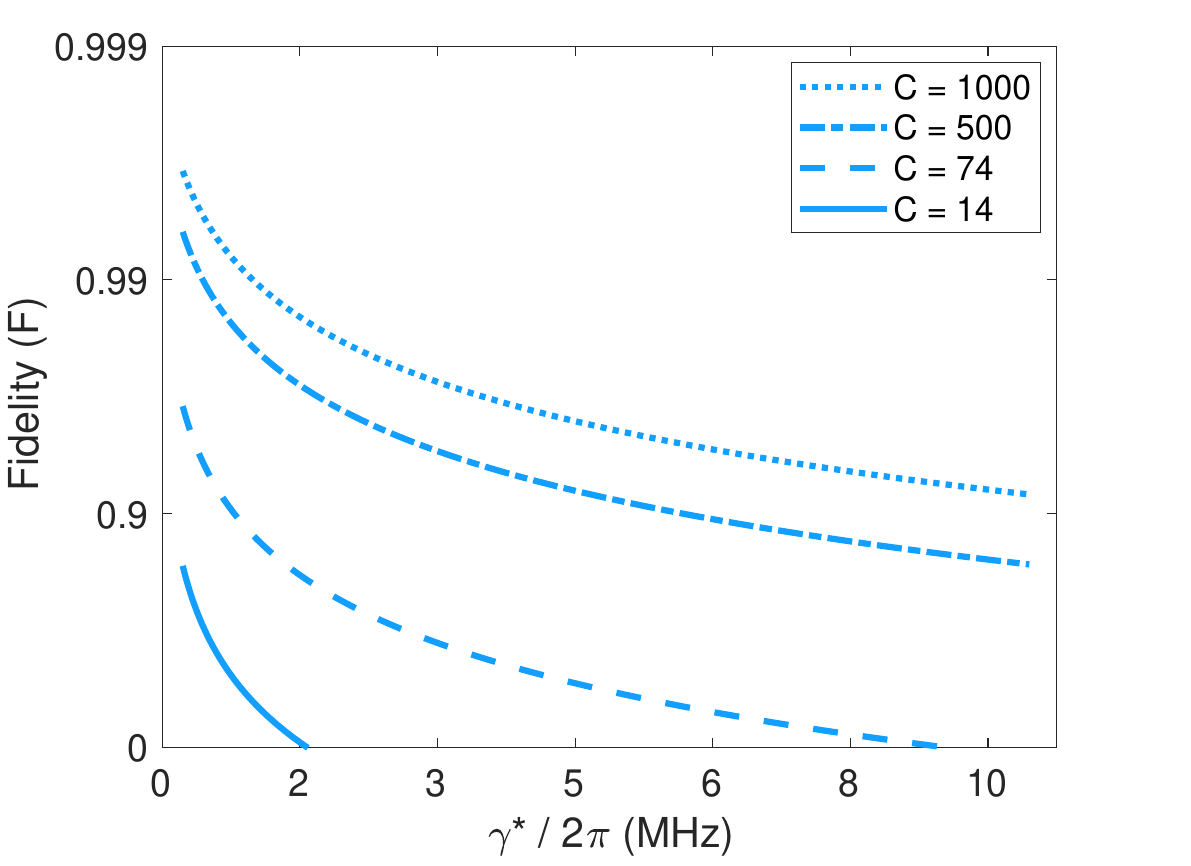}
\caption{Fidelity of the SB scheme as a function of the optical dephasing rate ($\gamma^*$) for different values of cavity cooperativity ($C$). The $\sigma_p$ value is optimized for each $C$ to maximize the fidelity. Note that the cooperativity values $C = 14$ and $C = 74$ correspond to the two scenarios shown in figure \ref{FetavsTd}.}
\label{FSBVSOD}
\end{figure}
The fidelity of the gate can be calculated by analyzing the interaction between a single photon and the cavity–qubit system using quantum Langevin equations, in combination with numerical methods \cite{PhysRevA.94.043807, asadi2020cavity, walls2008quantum}. The fidelity of performing the SB scheme is well approximated by \cite{asadi2020cavity, Faezehcavityerratum}
\begin{multline}
F_{\text{SB}} = 1 - \left( \frac{11\gamma^*}{8\gamma C} \right) - \frac{11}{16 C^2}- \Gamma T_{\text{SB}}  \\
- \frac{\left(-11 + 10 \left(\frac{2g}{\kappa}\right)^2\right) \, \delta_p \, (\delta_{\epsilon_{A}}+\delta_{\epsilon_{B}})}{4 \gamma^2 C^2} \\
- \frac{41 \delta_{\epsilon_{A}}^2 - 38 \delta_{\epsilon_{A}} \delta_{\epsilon_{B}} + 41 \delta_{\epsilon_{B}}^2}{16 \gamma^2 C^2} \\
- \frac{\left(11 - 20 \left(\frac{2g}{\kappa}\right)^2 + 12 \left(\frac{2g}{\kappa}\right)^4\right)(\delta_p^2 + \sigma_p^2)}{4 \gamma^2 C^2}
\label{fidelity scattering}    
\end{multline}
where $g$ represents the cavity coupling strength, $\kappa$ is the decay rate of the cavity, $\delta_p$ is the mean cavity-photon detuning, and $\Gamma = 1/T_{2e}$ is the effective decoherence rate for the ground state (the electron spin states). 
\\\indent As expected, increasing the cooperativity $C$ while keeping other parameters constant results in higher fidelities. In contrast, increasing $\delta_p$, $\delta_{\epsilon_{A}}$, $\delta_{\epsilon_{B}}$ (which can arise from spectral diffusion in the system), or the ratio $\frac{\gamma^*}{\gamma}$ leads to a reduction in fidelity. Considering that $T_{\mathit{SB}}\propto 1/\sigma_p$, there is a trade-off between the errors arising from the fourth and seventh terms. By optimization of the fidelity with respect to $\sigma_p$ one can see what value of the spectral standard deviation of the photon $\sigma_p$ leads to the maximum fidelity for each cavity cooperativity $C$. The optimized $\sigma_p$ is
\begin{equation}
\sigma_p = 2\,\sqrt[3]{\frac{4 \pi\sqrt{2\ln{2}} \,\Gamma \gamma^2 C^2 }{\left[11 - 20 \left(\frac{2g}{\kappa}\right)^2 + 12\left(\frac{2g}{\kappa}\right)^4\right]}}.  
\label{sigmap}
\end{equation}
Using the experimental values $g = 2\pi \times 42.4 \, \mathrm{MHz}$ and $\kappa = 2\pi \times 5.22 \, \mathrm{GHz}$ from \cite{johnston2024cavity}, which yield a ratio of $g/\kappa = 0.008$, and assuming zero detuning between T centers and the cavity mode ($\delta_{\epsilon_{A}} = \delta_{\epsilon_{B}} = 0$), figure \ref{FSBVSOD} illustrates the fidelity of the SB gate as a function of the $\gamma^*$, evaluated for various cavity cooperativity values ($C$). For each $C$, the value of $\sigma_p$ is optimized to achieve the maximum fidelity. The considered $\gamma^*$ values for T centers span from the values achieved in bulk \cite{higginbottom2023memory} to those measured in device-integrated implementations \cite{afzal2024distributed}. As expected, increasing $\gamma^*$ leads to a decrease in fidelity, highlighting the detrimental effect of optical dephasing on gate performance. Higher cooperativity values consistently result in higher fidelities across the considered range of $\gamma^*$. For example, with $C = 1000$, the fidelity remains above 0.99 even at moderate dephasing rates, whereas for $C = 14$, it drops sharply and becomes very low at the same $\gamma^*$ values.
\\\indent We explored two additional cavity-based schemes: the simple virtual photon exchange scheme and the Raman virtual photon exchange scheme, with details provided in the Supplementary Material (see sections \ref{simple} and \ref{Raman}). However, these schemes appear to be less promising for T centers in the near and medium term compared to the SB scheme.

\section{deterministic magnetic dipole-based scheme}
\label{MDG}
By leveraging the magnetic dipolar interaction between two nearby T centers, a two-qubit quantum gate can be implemented. We adapt the approach from Ref. \cite{grimm2021universal}, originally proposed for rare-earth ions, to T centers based on the analysis in Ref. \cite{karimi2024comparing}. In this proposal, the ``passive'' qubits store quantum information, while the ``active'' qubits, which possess a magnetic moment, interact magnetically to perform a two-qubit phase gate \cite{grimm2021universal}. We begin by initializing the T centers in the lowest hyperfine level of the ground state, then apply a $\pi/2$ microwave pulse to create a superposition of the states (passive qubits). The passive qubits are activated by applying two $\pi$ pulses to each system, transferring them to the active qubits. An Ising-type spin-spin interaction between the active qubits naturally introduces a phase shift, based on their states, through the unitary time evolution operator $U_{\mathit{Ising}} = e^{i(\pi/4) \sigma_Z^{1} \sigma_Z^{2}}$ \cite{vandersypen2004nmr}. Subsequently, the active qubits are returned to the passive qubits by applying another set of $\pi$ pulses.
By applying the Ising-type interaction in combination with single-qubit gates, one can realize a CZ gate \cite{karimi2024comparing}, a controlled-phase (CPhase) gate \cite{vandersypen2004nmr}, or a CNOT gate \cite{vandersypen2004nmr, hill2015surface, grimm2021universal, karimi2024comparing}.
\begin{figure}[h!]
\centering
\includegraphics[scale=0.45]{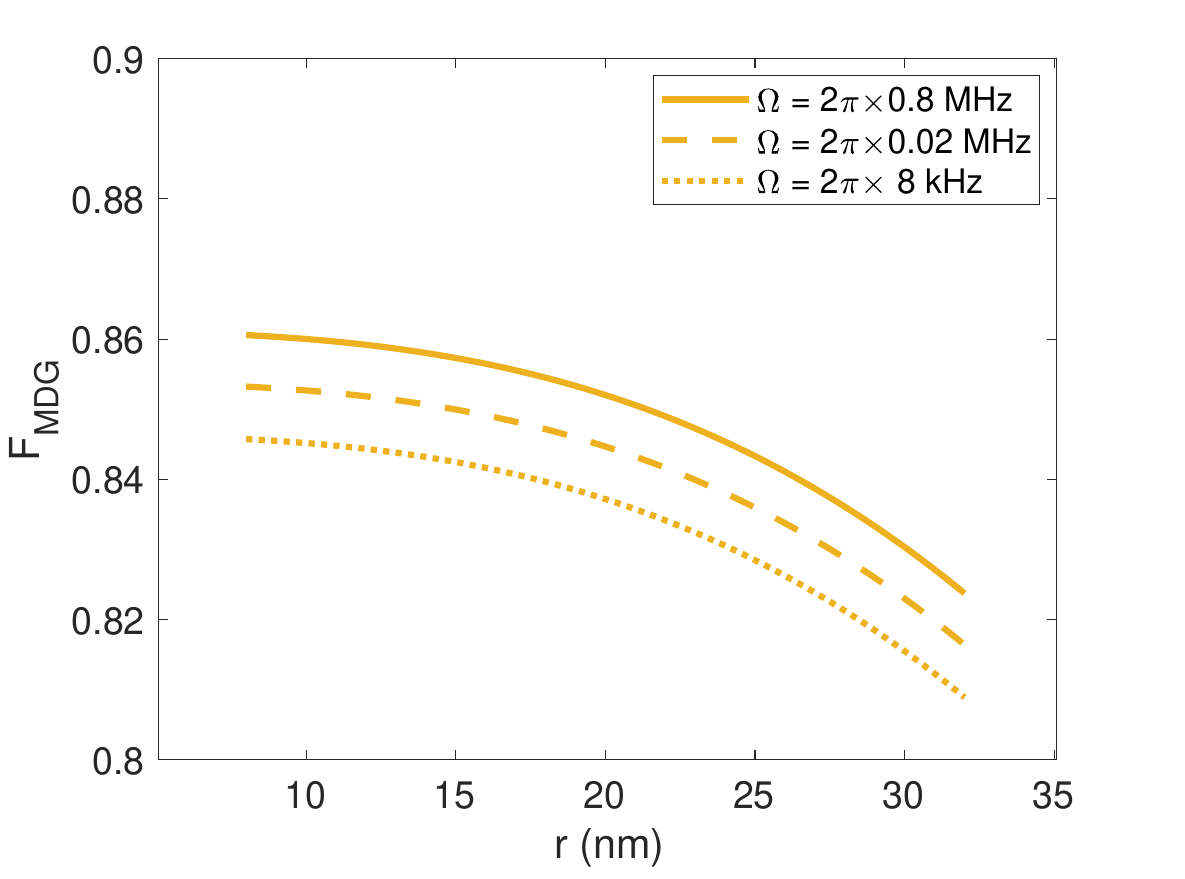}
\caption{Fidelity of performing the magnetic dipole gate between ground states as a function of the distance ($r$) between two T centers, shown for different Rabi frequencies ($\Omega$) affecting the infidelity from unwanted detuned transitions (see text for more detail).}
\label{FMDGVSromega}
\end{figure}
The total gate time for the magnetic dipole gate is $T_{\mathit{MD}} = 2T_{act} + T_{int}$ where the first term $T_{act} = \pi/\Omega$ is the time to excite from passive to active qubits and bring them back to the passive qubits. $\Omega$ is the Rabi frequency for $\ket{0}_{passive}$ to $\ket{0}_{active}$ and $\ket{1}_{passive}$ to $\ket{1}_{active}$ transitions. The second term $T_{int} = \hbar \pi/4 J_{z}$ is the time to perform the Ising interaction between two active qubits in the excited states, with the dipolar coupling strength $J_{z} = \frac{\mu_0 (\mu_B g_{z})^2}{8 \pi r^3}$ \cite{grimm2021universal, karimi2024comparing}, where $\mu_0$ is the permeability of free space, $\mu_B$ is the electron Bohr magneton, and $g_{z}$ is the principal value of the g-tensor of the excited state (hole spin). Applying tools from time-dependent perturbation theory that includes solving the Schrödinger equation to determine the ideal gate evolution and calculating the lowest-order error expressions, the fidelity of the gate is \cite{karimi2024comparing}
\begin{multline}
F_{\text{MD}} = 1 - t_{act} [\frac{7}{8}(\gamma_{1\uparrow} + \gamma_{1\downarrow})+\frac{13}{16}(\gamma_2 + \gamma_3)+\\ \frac{1}{2}(\gamma_4+\gamma_5)]  
 - t_{int}[\gamma_{1\uparrow}+\gamma_{1\downarrow}+\frac{3}{4}\gamma_3 +\frac{1}{2}\gamma_5]- a\frac{(J_{x}+J_{y})^2}{J_{z}^2}
\label{FMD}
\end{multline}
where $\gamma_{1\uparrow}(\gamma_{1\downarrow})$ is the decay rate for transition $\ket{0}_{active}$ to $\ket{0}_{passive}$ ($\ket{1}_{active}$ to $\ket{1}_{passive}$), $\gamma_2$ ($\gamma_3$) is the spin decay rate of the passive (active) qubits, $\gamma_4$ ($\gamma_5$) is the spin dephasing rate of the passive (active) qubits. There is a second order term in equation \eqref{FMD} with the coefficient $a = [32\,(2-\sqrt2)-\pi^2]/64$ and $J_{\alpha} = \frac{\mu_0 (\mu_B g_{\alpha})^2}{16 \pi r^3}$ $(\alpha = x, y)$ being the transverse components of the Ising-based interaction \cite{karimi2024comparing}. Assuming $g_x = g_y \equiv g_{\perp}$ and $g_z \equiv g_{\parallel}$, we investigate two choices of passive and active qubits, evaluating the resulting gate time and fidelity. 
\\\indent Here, we present the results for performing the magnetic dipole gate in the ground state levels (MDG) of the T center. In figure \ref{Tcenter}\,\textbf{b}, levels $\ket{1}$ and $\ket{2}$ are chosen as passive qubits, while levels $\ket{1}$ and $\ket{4}$ are the active qubits. Thus, level $\ket{1}$ is common to both the passive and active qubit definitions. To have a strong enough magnetic dipole interaction the two excited states should have different electron spins as the nuclear magneton $\mu_{N}$ is much smaller compared to the Bohr magneton $\mu_{B}$. 
\\\indent The electron spin $g_e$ tensor is treated as isotropic here, although experimental evidence suggests a minor anisotropy at the $10^{-3}$ level \cite{clear2024optical}. Assuming $g_x = g_y = g_z = 2.01$ \cite{clear2024optical, bergeron2020silicon}, and a branching ratio of $90\%$ for the $\mathrm{MW_{\Downarrow}}$ transition and $10\%$ for the forbidden $\mathrm{X_{m}}$ transition, figure \ref{FMDGVSromega} shows the fidelity $F_{\mathit{MDG}}$ of implementing the magnetic dipole gate between ground-state levels as a function of the distance $r$ between two T centers and for different Rabi frequencies ($\Omega$). The applied Rabi frequency $\Omega$ should be smaller than the splitting between the nuclear spins of the $\ket{\downarrow_e}$ to ensure the correct level is addressed. The authors in Ref. \cite{grimm2021universal} quantified the infidelity arising from unwanted detuned transitions while performing $\pi$ pulses as $(\frac{\pi}{2})^2 \exp[-\frac{\pi}{2}(\frac{\Delta_n}{\Omega})^2]$. For a splitting of $\Delta_n = 2\pi\times 2.1 \, \mathrm{MHz}$ \cite{bergeron2019optical}, choosing $\Delta_n/\Omega \approx 3$ results in an error of approximately $10^{-6}$ in the fidelity. The fidelity decreases with increasing the separation $r$ due to the weakening of the magnetic dipolar coupling strength over distance. It is moderately sensitive to variations in $\Omega$, with higher Rabi frequencies resulting in an improved gate performance. This highlights the importance of stronger driving fields in compensating for the reduced interaction strength at larger distances.
\\\indent We investigated two additional deterministic dipole-based schemes: the electric dipole-based gate and the magnetic dipole-based gate in the excited state with the details provided in the Supplementary Material (see sections \ref{elecgate} and \ref{MDE}).  However, these schemes seem to be less promising for T centers in the near and medium term. 
\\\indent Although performing the gate in the ground state is relatively slow compared to the excited state (see figure \ref{FTdistancedep}), the total gate time is still much less than the electron spin lifetime of $16 \,\mathrm{s}$ and the decoherence time of $2.1 \,\mathrm{ms}$ showing the feasibility of performing the magnetic dipole gate in the T center ground state.

\section{Gate performance comparison}
\label{comparison}
\begin{figure}[h!]
\centering
\includegraphics[scale=0.42]{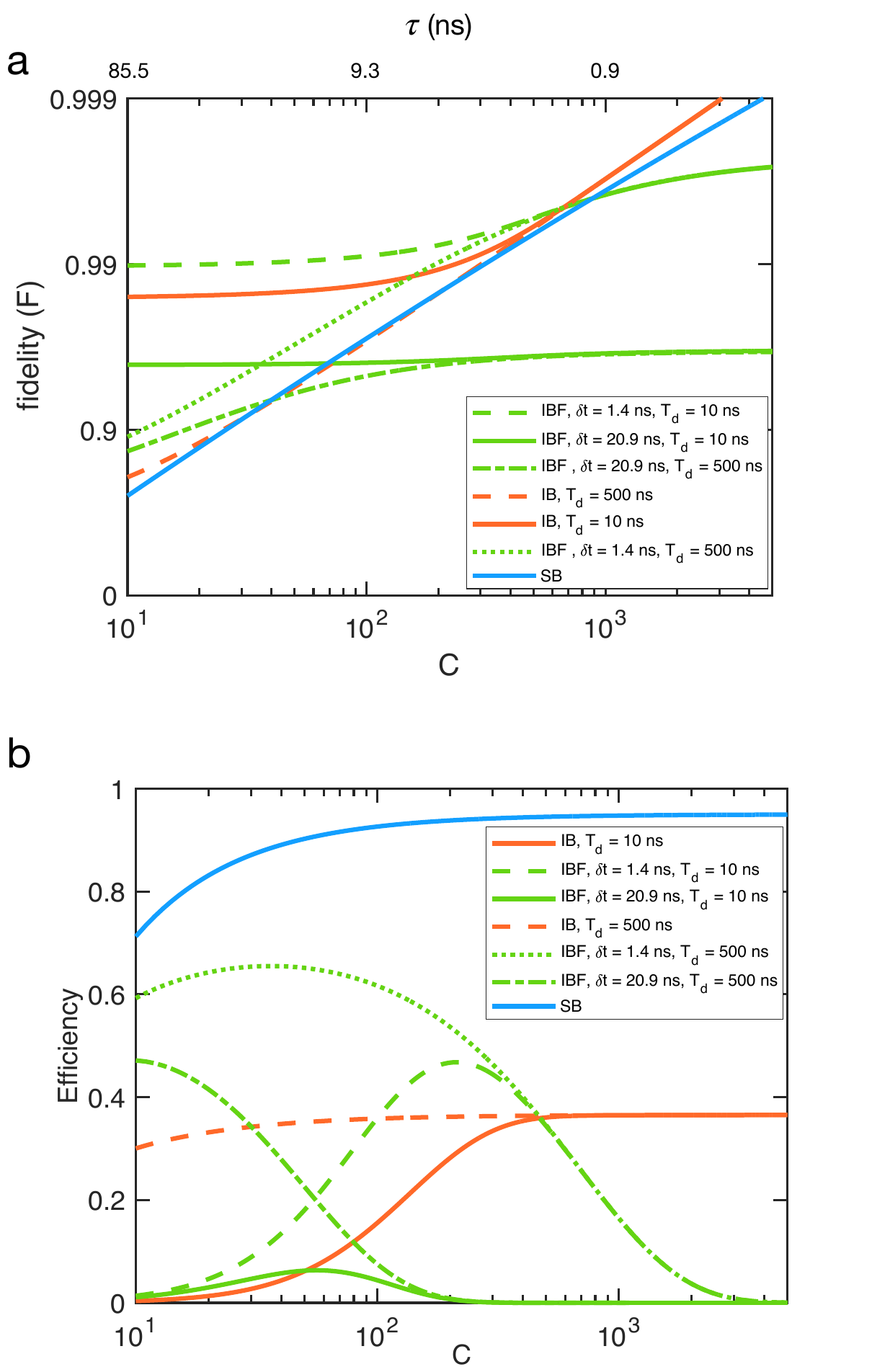}
\caption{A comparison of the \textbf{(a)} fidelity and \textbf{(b)} efficiency of the probabilistic and near-deterministic photon-mediated gates (IB, IBF, and SB), with respect to the cavity cooperativity $C$. The enhanced emitter lifetimes ($\tau$) corresponding to the values of $C$ are labeled along the top axis of figure \textbf{(a)}. The fidelity of the SB scheme is calculated using the optimized value of $\sigma_p$ for each $C$ value. The delay time values $\delta t$ correspond to the scenarios shown in figure \ref{FetavsTd}, but with varying $C$ (and therefore $\gamma'$ and $F_p$). For the IB and IBF schemes, the results are shown at two detection times, as presented in figure \ref{FetavsTd}.}
\label{FetavsC}
\end{figure}

In this section, we compare the performance of the probabilistic photon-mediated schemes (IB, IBF), the near-deterministic photon scattering-based gate (SB), and the deterministic magnetic dipole-based gate in the ground states (MDG) analyzed here. 
\\\indent First, we evaluate the dependence of the efficiency and fidelity of the probabilistic photon-mediated schemes on the cavity cooperativity $C$ (see figure \ref{FetavsC}\,\textbf{a} (\textbf{b})). $F_p = \frac{4g^2}{\kappa\gamma_{\mathit{zpl}}}$ \cite{janitz2020cavity} and $C = \frac{4g^2}{\kappa\gamma}$ \cite{o2016nondestructive, asadi2020protocols} result in $F_p = \frac{C}{\eta_r\, \eta_{zpl}}$, and $\gamma'$ is given by $\gamma' = \gamma \, (1+C)$.

\begin{figure}[h!]
\centering
\includegraphics[scale=0.43]{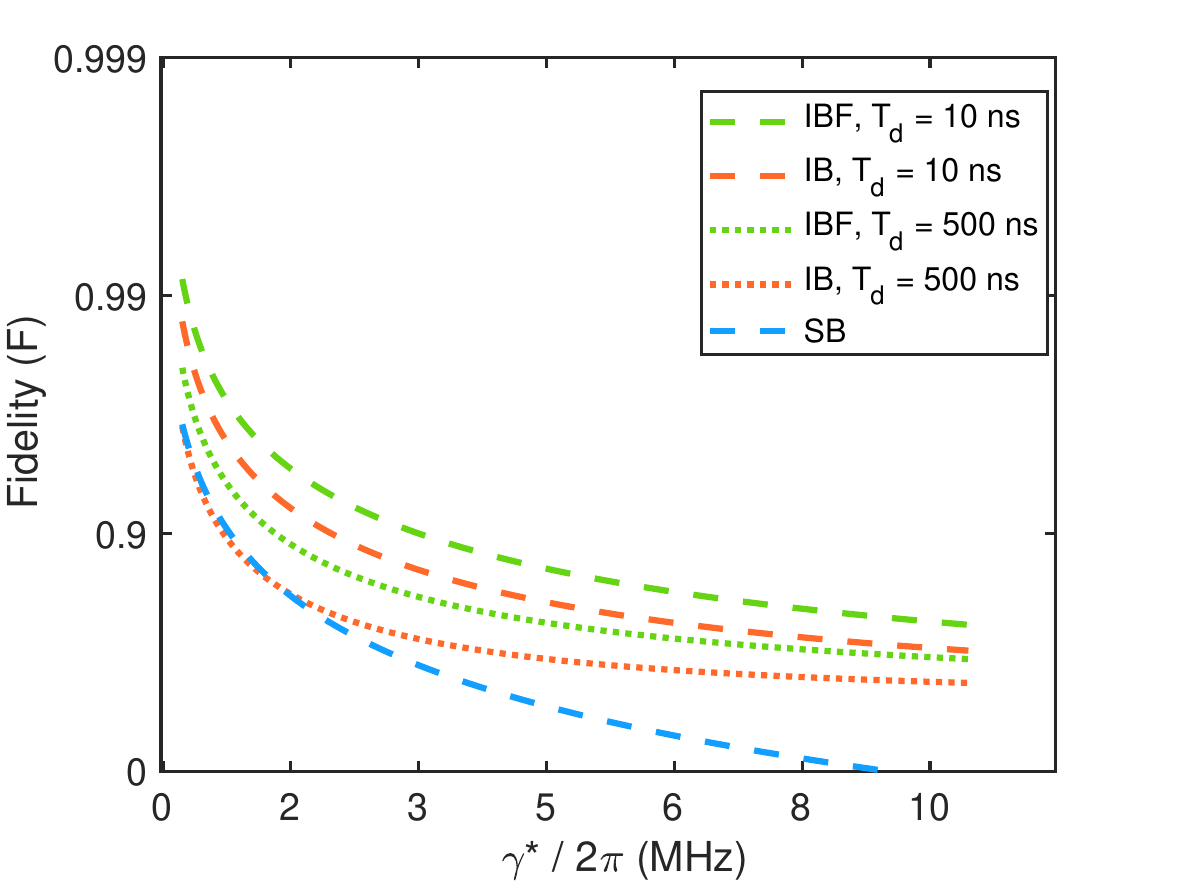}
\caption{A comparison of the fidelity of probabilistic and near-deterministic photon-mediated gates (IB, IBF, and SB) as a function of the optical dephasing rate ($\gamma^*$). The $\gamma^*$ range spans from the best achievable value, found in bulk, to the value observed in device. For all the mentioned schemes, scenario 2 is shown, which means that for the IB and IBF schemes, the parameters used are the enhanced optical decay rate $\gamma' = 2\pi\times12.7 \, \mathrm{MHz}$, Purcell factor $F_p = 1402.4$, and delay time $\delta t = 1.4\,\mathrm{ns}$ (see figure \ref{FetavsTd}). For the SB scheme, the cooperativity value $C = 74$ is used, corresponding to the $\gamma'$ value of the scenario 2. For the IB and IBF schemes, the results are shown at two detection times, as presented in figure \ref{FetavsTd}.}
\label{compareALL}
\end{figure}

Following the discussion in section \ref{probabilistic photon-mediated gates}, we evaluate the efficiency and fidelity of the IB and IBF schemes at two detection times: $T_{d} = 10\, \mathrm{ns}$ and $T_{d} = 500 \,\mathrm{ns}$, which corresponds to the trade-off between efficiency and fidelity (see figure \ref{FetavsTd}). While the efficiency improves with $C$ for the IB and SB schemes, this is not the case for the IBF scheme, where $\eta_{\mathit{IBF}}=0$ at sufficiently large $C$ when $\gamma'$ increases (see equation \eqref{etaIBF}). This occurs because the protocol fails if the second photon is detected before the feedback pulse is completed; thus, rapid decay is undesirable. In the medium term, we focus on the low-to-moderate $C$ regime. For the same detection times ($T_{d}$), the IBF scheme outperforms the IB scheme in efficiency at low cooperativities ($C < 50$) across both scenarios. In the moderate cooperativity regime, the efficiency of IBF with $\delta t = 1.4\, \mathrm{ns}$ exceeds that of the IB scheme. Note that it is still possible to achieve high efficiencies at large $C$ if delay times shorter than $\delta t = 1.4\, \mathrm{ns}$ can be realized, using techniques such as Raman transitions employed in \cite{jenkins2022ytterbium} to achieve fast Rabi oscillations. In practice, implementing feedforward requires more than just the spin-flip operation time; it also involves processing photon detection events and triggering pulses \cite{thiele2025cryogenic}. The SB gate provides the best efficiency scaling with $C$ compared to probabilistic schemes. However, at low and moderate $C$, the probabilistic IB and IBF schemes offer higher fidelities than the SB scheme. Among them, IBF $(\delta t = 1.4\, \mathrm{ns})$ at $T_{d} = 10 \,\mathrm{ns}$ yields the highest fidelity.
\\\indent Considering the impact of optical dephasing on the performance of the schemes, figure \ref{compareALL} shows a comparison of the fidelities of the IB, IBF, and SB schemes as a function of $\gamma^*$ under scenario 2 parameters. As the $\gamma^*$ increases, the fidelity of all schemes decreases that reflects the detrimental impact of optical dephasing. Among the schemes, IBF with a short detection time ($T_d = 10\, \mathrm{ns}$) achieves the highest fidelity and exhibits the greatest robustness against optical dephasing. In contrast, the SB scheme shows the lowest fidelities across the entire range of $\gamma^*$. For both IB and IBF, increasing the detection time from $10\, \mathrm{ns}$ to $500\, \mathrm{ns}$ leads to a notable reduction in fidelity, consistent with the results shown in figure \ref{FetavsTd}, highlighting the importance of temporal filtering enabled by fast detection in mitigating optical dephasing effect \cite{ngan2024performance}. 
\begin{figure}[h!]
\centering
\includegraphics[scale=0.42]{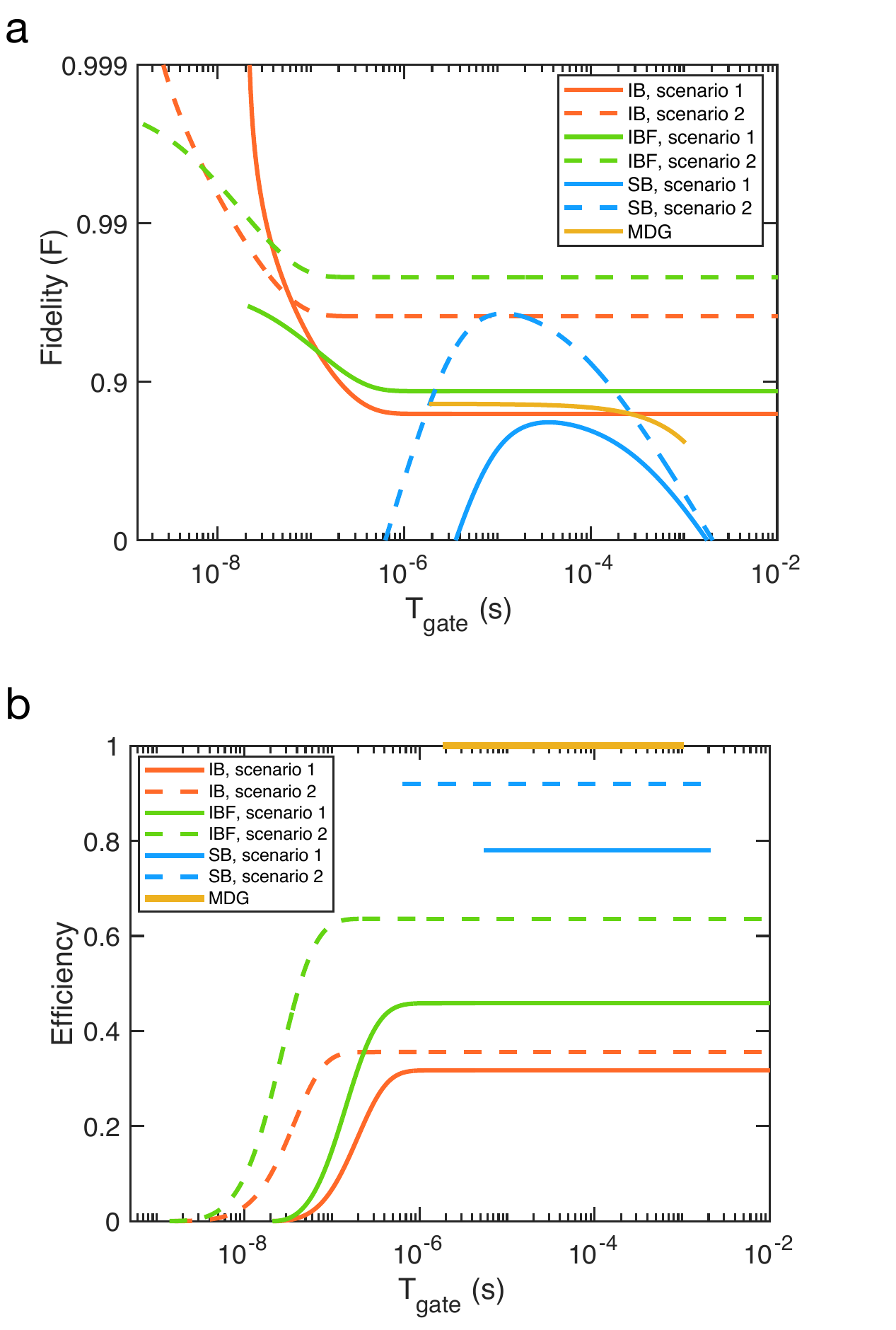}
\caption{Figure shows \textbf{(a)} the fidelity ($F$), and \textbf{(b)} efficiency with respect to the gate time $T_{gate}$ for all the gates in sections \ref{probabilistic photon-mediated gates}, \ref{SB}, and \ref{MDG}. Each color corresponds to a certain gate. Scenario 1 uses the currently demonstrated experimental values of the enhanced optical decay rate $\gamma' = 2\pi\times2.5 \, \mathrm{MHz}$, Purcell factor $F_p = 256.5$, and delay time $\delta t = 20.9 \, \mathrm{ns}$, whereas scenario 2 considers the more optimistic values of $\gamma'=2\pi\times12.7 \, \mathrm{MHz}$, $F_p=1402.4$, and $\delta t = 1.4\,\mathrm{ns}$ (see figure \ref{FetavsTd}). For the SB scheme, the change in fidelity and efficiency with respect to $T_{gate}$ is shown for two $C$ values to facilitate better evaluation. These values correspond to the $\gamma'$ values of the scenarios 1 and 2: $C = 14$ and $C= 74$. Optical dephasing is set to the bulk value in both scenarios \cite{higginbottom2023memory}. For the MDG scheme, both the fidelity and the gate time $T_{gate}$ are expressed as functions of the distance $r$, and $\Omega = 2\pi\times 0.8 \,\mathrm{MHz}$, which corresponds to the case shown in figure \ref{FMDGVSromega}.}
\label{FetavsTgate}
\end{figure}
\\\indent Figure \ref{FetavsTgate} \,\textbf{a} (\textbf{b}) shows the fidelity (efficiency) as a function of gate time $T_{gate}$ for all the gates investigated in sections \ref{probabilistic photon-mediated gates}, \ref{SB}, and \ref{MDG}. For the MDG scheme, fidelity and $T_{gate}$ are expressed as functions of the distance ($r$) between two T centers (see figure \ref{FTdistancedep} for corresponding plots). While MDG fidelity decreases with increasing gate time, SB fidelity initially rises before declining after reaching a maximum, reflecting a trade-off between the fourth and seventh terms in the fidelity equation (equation \eqref{fidelity scattering}). The fidelity of probabilistic photon-mediated gates follows the same trend as their dependence on detection time $T_{d}$, with a plateau at longer gate times (see section \ref{probabilistic photon-mediated gates} for the relationship between $T_{gate}$ and $T_{d}$, and figure \ref{FetavsTd} for a comparison). Excluding very short gate times, the IBF scheme consistently achieves higher fidelities than the IB scheme in each scenario.
\\\indent In terms of efficiency versus $T_{gate}$, the MDG scheme has the advantage of unit efficiency as a deterministic gate. For probabilistic photon-mediated gates, efficiency increases with gate time until it reaches a maximum and plateaus in the optical limit regime. Similar to fidelity, the IBF scheme achieves higher efficiency than the IB scheme in each scenario. As noted in section \ref{probabilistic photon-mediated gates}, interference-based schemes exhibit a trade-off: fidelity is higher at shorter times, whereas efficiency improves with longer times.
\\\indent Our findings suggest that, given near and medium-term design and fabrication capabilities, the probabilistic interference-based schemes are the most promising for implementation in T centers. The near-deterministic SB scheme is also favorable at higher cooperativities $C$ or when prioritizing efficiency over fidelity. The MDG scheme offers a high fidelity and deterministic operation, albeit with a slower gate time, if T centers can be positioned in close proximity, similar to other proximity-based gates. 
\section{conclusion and outlook}
\label{conclusion}
In summary, we have investigated various gates between individual T centers in silicon, analyzing and comparing two probabilistic photon interference-based schemes (IB and IBF), a near-deterministic photon scattering-based (SB) scheme, and a deterministic magnetic dipole-based scheme in the ground state (MDG), given current and near-future experimental and technological advancements. The schemes were evaluated using various metrics, including fidelity, efficiency, and gate time, while importantly accounting for practical imperfections in the system and experimental setup. 
\\\indent To enable a meaningful comparison between schemes, we performed a detailed analysis of the photon interference-based scheme with feedback (IBF) and, for the first time, quantified its efficiency and fidelity. We derived new analytical expressions that account for real-world imperfections, including optical decoherence, spin dephasing, photon loss, and inefficiencies in photon collection and detection.
\\\indent Considering the fabrication technologies for cavities and the creation of T centers in the near and medium term, we conclude that implementing probabilistic photon interference-based schemes is promising. In the near term, while the IB scheme has already been demonstrated \cite{afzal2024distributed}, the IBF scheme appears particularly attractive, as it simultaneously achieves the highest efficiency and fidelity at low (as observed in current experiments \cite{johnston2024cavity, islam2023cavity}) and moderate values of $C$, making it a strong candidate for experimental implementation. In the long term, we infer the feasibility of implementing the MDG and SB schemes between T centers. 
\\\indent Considering the requirements for feedforward implementation, cryogenic complementary metal-oxide-semiconductor (cryo-CMOS) technologies have the potential to achieve feedback delays below $10\,\mathrm{ns}$—a threshold that currently represents the experimental limit. However, realizing such low-latency control remains challenging experimentally. The potential for higher efficiency with the IBF scheme further motivates the development of feedback systems capable of operating within this sub-$10\,\mathrm{ns}$ regime.
\\\indent The results in this paper assume that spectral diffusion (SD) can be controlled, though mitigating its impact remains challenging. Several works have addressed this issue \cite{metz2008effect, kambs2018limitations, uysal2024rephasing}, including time-filtering techniques \cite{afzal2024distributed} and resonance-check methods \cite{bowness2025laser, zhang2025laser}, which have been shown to reduce linewidth by up to 35× in T center devices \cite{bowness2025laser}. However, intrinsic spectral mismatch and almost lifetime-limited homogeneous linewidths constrain the achievable spectral alignment between T centers.  Active tuning strategies, such as electric field tuning, have also been explored to mitigate these mismatches \cite{clear2024optical}. Moreover, SD in T centers has been observed to depend on excitation power, with entanglement pulses capable of inducing spectral shifts \cite{bowness2025laser}. This suggests schemes less dependent on excitation lasers may be more resilient to SD. To account for these effects in modeling, a common approach is to average fidelity over a bivariate Gaussian distribution representing emission fluctuations \cite{wein2020analyzing, ngan2024performance}. While this approach captures the effects of spectral diffusion statistically, it does not account for the excitation-power-dependent nature of the shifts. A more refined approach for modeling can be to consider SD as a time-dependent function, relating fidelity to pulse sequence characteristics and excitation-induced frequency shifts.
\\\indent It is worth noting that the efficiency presented here is based on an absolute time-bin filter, in which only photons detected within a fixed time window after excitation are considered. While this approach can be advantageous—since spectral diffusion has had minimal time to affect the emitter's frequency during early emission—it may underestimate the achievable efficiency compared to the more commonly used correlation filter \cite{afzal2024distributed}. In a correlation filter, photons are selected based on their relative detection times; that is, they are accepted if their arrival times differ by less than a specified time window. This method captures additional successful events, such as photons emitted at later times but still within the accepted temporal correlation, so increasing the overall success probability. Although correlation filtering might not be immediately compatible with the IBF scheme due to the feedforward control, it remains an open question whether a similar filtering strategy could be adapted to IBF, which is worth investigating experimentally. Nevertheless, in the short-lifetime regime—where photon indistinguishability is highest—the difference between these filtering methods becomes less significant, and absolute filtering may even be advantageous due to suppression of the effects of SD by favoring early emission events.
\\\indent Given the promising outcomes of the probabilistic photon-mediated interference-based schemes for single T centers, one could also investigate the feasibility of other probabilistic interference-based schemes, such as single-photon detection and polarization-based schemes \cite{sangouard2011quantum, wein2020analyzing, lago2021telecom}.
\\\indent Since the analytical equations for efficiency and fidelity obtained here can be applied to other systems, and given that the IBF scheme achieves competitive fidelity and efficiency compared to the IB scheme in T centers, one can explore the implementation of this scheme and compare the performance of the schemes discussed here—and in the Supplementary Material—across other solid-state quantum systems.  These include various color centers such as those found in diamond (e.g., NV and Si-V centers), as well as rare-earth-ion-doped crystals. Inherent system properties, such as optical decay and decoherence rates, as well as technological considerations—including the feasibility of system creation, spatial localization of defects, and achievable cavity coupling—would be key factors in determining the most promising gates for each system.
\\\indent Because the IBF scheme is sensitive to the feedback delay time $\delta t$, it is best suited for performing local gates, i.e., between modules in close physical proximity \cite{martin2019single}, rather than for communication over extended distances (specifically in T centers). However, depending on the physical system employed, the requirement for short feedback delay times can be relaxed, reducing the difficulty of implementing feedforward control. For example, $\mathrm{Er^{3+}}:\mathrm{Y}_2\mathrm{SiO}_5$ is a promising solid-state platform with emission in the telecommunication band and a millisecond-scale excited-state lifetime (on the order of $10\,\mathrm{ms}$) \cite{mcauslan2009strong}, allowing larger values of $\delta t$ and enabling feedback between more distant nodes with looser constraints on feedforward control. On the other hand, shorter feedback delay times would be easier to achieve in room-temperature platforms, which don't suffer from the limitations imposed by signal propagation through cryostat wiring and limited cooling capacity.
\\\indent Other potential next steps could include investigating the feasibility of implementing quantum repeaters with single T centers, as has already been explored for rare-earth ions \cite{asadi2018quantum, asadi2020protocols} and NV centers \cite{ji2022proposal, rozpkedek2019near}. Additionally, the potential for distributed quantum computing (DQC) based on these various gates could be investigated \cite{main2025distributed, barral2024review}. In particular, DQC has been explored using the IB scheme in recent work \cite{afzal2024distributed}, discussed in roadmaps \cite{simmons2024scalable}, and remains an active area of research.

\begin{acknowledgments}
We wish to thank Mahsa Karimi for useful discussions and acknowledge support from the NSERC Alliance Quantum Consortia Grants ARAQNE and QUINT, and the NRC CSTIP Challenge Program on High-Throughput Secure Networks (HTSN).
\end{acknowledgments}

\medskip
\bibliographystyle{unsrt}
\bibliography{references}

\begin{thebibliography}{10}

\bibitem{nielsen2010quantum}
Michael~A Nielsen and Isaac~L Chuang.
\newblock {\em Quantum computation and quantum information}.
\newblock Cambridge university press, 2010.

\bibitem{deutsch1989quantum}
David~Elieser Deutsch.
\newblock Quantum computational networks.
\newblock {\em Proceedings of the royal society of London. A. mathematical and physical sciences}, 425(1868):73--90, 1989.

\bibitem{ladd2010quantum}
Thaddeus~D Ladd, Fedor Jelezko, Raymond Laflamme, Yasunobu Nakamura, Christopher Monroe, and Jeremy~Lloyd O’Brien.
\newblock Quantum computers.
\newblock {\em nature}, 464(7285):45--53, 2010.

\bibitem{sangouard2011quantum}
Nicolas Sangouard, Christoph Simon, Hugues De~Riedmatten, and Nicolas Gisin.
\newblock Quantum repeaters based on atomic ensembles and linear optics.
\newblock {\em Reviews of Modern Physics}, 83(1):33--80, 2011.

\bibitem{duan2001long}
L-M Duan, Mikhail~D Lukin, J~Ignacio Cirac, and Peter Zoller.
\newblock Long-distance quantum communication with atomic ensembles and linear optics.
\newblock {\em Nature}, 414(6862):413--418, 2001.

\bibitem{simon2017towards}
Christoph Simon.
\newblock Towards a global quantum network.
\newblock {\em Nature Photonics}, 11(11):678--680, 2017.

\bibitem{kimble2008quantum}
H~Jeff Kimble.
\newblock The quantum internet.
\newblock {\em Nature}, 453(7198):1023--1030, 2008.

\bibitem{wehner2018quantum}
Stephanie Wehner, David Elkouss, and Ronald Hanson.
\newblock Quantum internet: A vision for the road ahead.
\newblock {\em Science}, 362(6412):eaam9288, 2018.

\bibitem{ruf2021quantum}
Maximilian Ruf, Noel~H Wan, Hyeongrak Choi, Dirk Englund, and Ronald Hanson.
\newblock Quantum networks based on color centers in diamond.
\newblock {\em Journal of Applied Physics}, 130(7), 2021.

\bibitem{pompili2021realization}
Matteo Pompili, Sophie~LN Hermans, Simon Baier, Hans~KC Beukers, Peter~C Humphreys, Raymond~N Schouten, Raymond~FL Vermeulen, Marijn~J Tiggelman, Laura dos Santos~Martins, Bas Dirkse, et~al.
\newblock Realization of a multinode quantum network of remote solid-state qubits.
\newblock {\em Science}, 372(6539):259--264, 2021.

\bibitem{knaut2023entanglement}
Can~M Knaut, Aziza Suleymanzade, Yan-Cheng Wei, Daniel~R Assumpcao, Pieter-Jan Stas, Yan~Qi Huan, Bartholomeus Machielse, Erik~N Knall, Madison Sutula, Gefen Baranes, et~al.
\newblock Entanglement of nanophotonic quantum memory nodes in a telecommunication network.
\newblock {\em arXiv preprint arXiv:2310.01316}, 2023.

\bibitem{lei2023quantum}
Yisheng Lei, Faezeh Kimiaee~Asadi, Tian Zhong, Alex Kuzmich, Christoph Simon, and Mahdi Hosseini.
\newblock Quantum optical memory for entanglement distribution.
\newblock {\em Optica}, 10(11):1511--1528, 2023.

\bibitem{taherizadegan2024towards}
Shahrzad Taherizadegan, Jacob~H Davidson, Sourabh Kumar, Daniel Oblak, and Christoph Simon.
\newblock Towards a realistic model for cavity-enhanced atomic frequency comb quantum memories.
\newblock {\em Quantum Science and Technology}, 9(3):035049, 2024.

\bibitem{grimm2021universal}
Manuel Grimm, Adrian Beckert, Gabriel Aeppli, and Markus M{\"u}ller.
\newblock Universal quantum computing using electronuclear wavefunctions of rare-earth ions.
\newblock {\em Prx Quantum}, 2(1):010312, 2021.

\bibitem{ohlsson2002quantum}
Nicklas Ohlsson, R~Krishna Mohan, and Stefan Kr{\"o}ll.
\newblock Quantum computer hardware based on rare-earth-ion-doped inorganic crystals.
\newblock {\em Optics communications}, 201(1-3):71--77, 2002.

\bibitem{asadi2020protocols}
F~Kimiaee Asadi, SC~Wein, and C~Simon.
\newblock Protocols for long-distance quantum communication with single 167er ions.
\newblock {\em Quantum Science and Technology}, 5(4):045015, 2020.

\bibitem{bernien2013heralded}
Hannes Bernien, Bas Hensen, Wolfgang Pfaff, Gerwin Koolstra, Machiel~S Blok, Lucio Robledo, Tim~H Taminiau, Matthew Markham, Daniel~J Twitchen, Lilian Childress, et~al.
\newblock Heralded entanglement between solid-state qubits separated by three metres.
\newblock {\em Nature}, 497(7447):86--90, 2013.

\bibitem{evans2018photon}
Ruffin~E Evans, Mihir~K Bhaskar, Denis~D Sukachev, Christian~T Nguyen, Alp Sipahigil, Michael~J Burek, Bartholomeus Machielse, Grace~H Zhang, Alexander~S Zibrov, Edward Bielejec, et~al.
\newblock Photon-mediated interactions between quantum emitters in a diamond nanocavity.
\newblock {\em Science}, 362(6415):662--665, 2018.

\bibitem{day2022coherent}
Matthew~W Day, Kelsey~M Bates, Christopher~L Smallwood, Rachel~C Owen, Tim Schr{\"o}der, Edward Bielejec, Ronald Ulbricht, and Steven~T Cundiff.
\newblock Coherent interactions between silicon-vacancy centers in diamond.
\newblock {\em Physical Review Letters}, 128(20):203603, 2022.

\bibitem{safonov1996interstitial}
AN~Safonov, EC~Lightowlers, Gordon Davies, P~Leary, R~Jones, and Sven {\"O}berg.
\newblock Interstitial-carbon hydrogen interaction in silicon.
\newblock {\em Physical review letters}, 77(23):4812, 1996.

\bibitem{safonov1999photoluminescence}
AN~Safonov and EC~Lightowlers.
\newblock Photoluminescence characterisation of hydrogen-related centres in silicon.
\newblock {\em Materials Science and Engineering: B}, 58(1-2):39--47, 1999.

\bibitem{minaev1981thermally}
NS~Minaev and AV~Mudryi.
\newblock Thermally-induced defects in silicon containing oxygen and carbon.
\newblock {\em physica status solidi (a)}, 68(2):561--565, 1981.

\bibitem{irion1985defect}
E~Irion, N~Burger, K~Thonke, and R~Sauer.
\newblock The defect luminescence spectrum at 0.9351 ev in carbon-doped heat-treated or irradiated silicon.
\newblock {\em Journal of Physics C: Solid State Physics}, 18(26):5069, 1985.

\bibitem{safonov1993hydrogen}
AN~Safonov and Edward~C Lightowlers.
\newblock Hydrogen related optical centers in radiation damaged silicon.
\newblock In {\em Materials Science Forum}, volume 143, pages 903--908. Trans Tech Publ, 1993.

\bibitem{leary1998interaction}
P~Leary, R~Jones, and Sven {\"O}berg.
\newblock Interaction of hydrogen with substitutional and interstitial carbon defects in silicon.
\newblock {\em Physical Review B}, 57(7):3887, 1998.

\bibitem{simmons2024scalable}
Stephanie Simmons.
\newblock Scalable fault-tolerant quantum technologies with silicon color centers.
\newblock {\em PRX Quantum}, 5(1):010102, 2024.

\bibitem{bergeron2020silicon}
L~Bergeron, C~Chartrand, ATK Kurkjian, KJ~Morse, H~Riemann, NV~Abrosimov, P~Becker, H-J Pohl, MLW Thewalt, and S~Simmons.
\newblock Silicon-integrated telecommunications photon-spin interface.
\newblock {\em PRX Quantum}, 1(2):020301, 2020.

\bibitem{macquarrie2021generating}
ER~MacQuarrie, Camille Chartrand, DB~Higginbottom, KJ~Morse, VA~Karasyuk, Sjoerd Roorda, and Stephanie Simmons.
\newblock Generating t centres in photonic silicon-on-insulator material by ion implantation.
\newblock {\em New Journal of Physics}, 23(10):103008, 2021.

\bibitem{higginbottom2022optical}
Daniel~B Higginbottom, Alexander~TK Kurkjian, Camille Chartrand, Moein Kazemi, Nicholas~A Brunelle, Evan~R MacQuarrie, James~R Klein, Nicholas~R Lee-Hone, Jakub Stacho, Myles Ruether, et~al.
\newblock Optical observation of single spins in silicon.
\newblock {\em Nature}, 607(7918):266--270, 2022.

\bibitem{deabreu2023waveguide}
Adam DeAbreu, Camille Bowness, Amirhossein Alizadeh, Camille Chartrand, NA~Brunelle, ER~MacQuarrie, NR~Lee-Hone, Myles Ruether, Moein Kazemi, ATK Kurkjian, et~al.
\newblock Waveguide-integrated silicon t centres.
\newblock {\em Optics Express}, 31(9):15045--15057, 2023.

\bibitem{komza2025multiplexed}
Lukasz Komza, Xueyue Zhang, Hanbin Song, Yu-Lung Tang, Xin Wei, and Alp Sipahigil.
\newblock Multiplexed color centers in a silicon photonic cavity array.
\newblock {\em Optica}, 12(9):1400--1405, 2025.

\bibitem{higginbottom2023memory}
Daniel~B Higginbottom, Faezeh~Kimiaee Asadi, Camille Chartrand, Jia-Wei Ji, Laurent Bergeron, Michael~LW Thewalt, Christoph Simon, and Stephanie Simmons.
\newblock Memory and transduction prospects for silicon t center devices.
\newblock {\em PRX Quantum}, 4(2):020308, 2023.

\bibitem{PhysRevLett.77.4812}
A.~N. Safonov, E.~C. Lightowlers, Gordon Davies, P.~Leary, R.~Jones, and S.~\"Oberg.
\newblock Interstitial-carbon hydrogen interaction in silicon.
\newblock {\em Phys. Rev. Lett.}, 77:4812--4815, Dec 1996.

\bibitem{dhaliah2022first}
Diana Dhaliah, Yihuang Xiong, Alp Sipahigil, Sin{\'e}ad~M Griffin, and Geoffroy Hautier.
\newblock First-principles study of the t center in silicon.
\newblock {\em Physical Review Materials}, 6(5):L053201, 2022.

\bibitem{afzal2024distributed}
Francis Afzal, Mohsen Akhlaghi, Stefanie~J Beale, Olinka Bedroya, Kristin Bell, Laurent Bergeron, Kent Bonsma-Fisher, Polina Bychkova, Zachary~ME Chaisson, Camille Chartrand, et~al.
\newblock Distributed quantum computing in silicon.
\newblock {\em arXiv preprint arXiv:2406.01704}, 2024.

\bibitem{barrett2005efficient}
Sean~D Barrett and Pieter Kok.
\newblock Efficient high-fidelity quantum computation using matter qubits and linear optics.
\newblock {\em Physical Review A}, 71(6):060310, 2005.

\bibitem{martin2019single}
Leigh~S Martin and K~Birgitta Whaley.
\newblock Single-shot deterministic entanglement between non-interacting systems with linear optics.
\newblock {\em arXiv preprint arXiv:1912.00067}, 2019.

\bibitem{wein2020analyzing}
Stephen~C Wein, Jia-Wei Ji, Yu-Feng Wu, Faezeh Kimiaee~Asadi, Roohollah Ghobadi, and Christoph Simon.
\newblock Analyzing photon-count heralded entanglement generation between solid-state spin qubits by decomposing the master-equation dynamics.
\newblock {\em Physical Review A}, 102(3):033701, 2020.

\bibitem{asadi2020cavity}
F~Kimiaee Asadi, SC~Wein, and Christoph Simon.
\newblock Cavity-assisted controlled phase-flip gates.
\newblock {\em Physical Review A}, 102(1):013703, 2020.

\bibitem{karimi2024comparing}
Mahsa Karimi, Faezeh~Kimiaee Asadi, Stephen~C Wein, and Christoph Simon.
\newblock Comparing the performance of practical two-qubit gates for individual \textsuperscript{171}yb ions in yttrium orthovanadate.
\newblock {\em arXiv preprint arXiv:2410.23613}, 2024.

\bibitem{lim2005repeat}
Yuan~Liang Lim, Almut Beige, and Leong~Chuan Kwek.
\newblock Repeat-until-success linear optics distributed quantum computing.
\newblock {\em Physical review letters}, 95(3):030505, 2005.

\bibitem{moehring2007entanglement}
David~L Moehring, Peter Maunz, Steve Olmschenk, Kelly~C Younge, Dzmitry~N Matsukevich, L-M Duan, and Christopher Monroe.
\newblock Entanglement of single-atom quantum bits at a distance.
\newblock {\em Nature}, 449(7158):68--71, 2007.

\bibitem{ruskuc2024scalable}
Andrei Ruskuc, Chun-Ju Wu, Emanuel Green, Sophie~LN Hermans, Joonhee Choi, and Andrei Faraon.
\newblock Scalable multipartite entanglement of remote rare-earth ion qubits.
\newblock {\em arXiv preprint arXiv:2402.16224}, 2024.

\bibitem{grice2011arbitrarily}
Warren~P Grice.
\newblock Arbitrarily complete bell-state measurement using only linear optical elements.
\newblock {\em Physical Review A—Atomic, Molecular, and Optical Physics}, 84(4):042331, 2011.

\bibitem{ewert20143}
Fabian Ewert and Peter van Loock.
\newblock 3/4-efficient bell measurement with passive linear optics and unentangled ancillae.
\newblock {\em Physical review letters}, 113(14):140403, 2014.

\bibitem{wein2016efficiency}
Stephen Wein, Khabat Heshami, Christopher~A Fuchs, Hari Krovi, Zachary Dutton, Wolfgang Tittel, and Christoph Simon.
\newblock Efficiency of an enhanced linear optical bell-state measurement scheme with realistic imperfections.
\newblock {\em Physical Review A}, 94(3):032332, 2016.

\bibitem{bayerbach2023bell}
Matthias~J Bayerbach, Simone~E D’Aurelio, Peter van Loock, and Stefanie Barz.
\newblock Bell-state measurement exceeding 50\% success probability with linear optics.
\newblock {\em Science Advances}, 9(32):eadf4080, 2023.

\bibitem{johnston2024cavity}
Adam Johnston, Ulises Felix-Rendon, Yu-En Wong, and Songtao Chen.
\newblock Cavity-coupled telecom atomic source in silicon.
\newblock {\em Nature Communications}, 15(1):2350, 2024.

\bibitem{grange2015cavity}
Thomas Grange, Gaston Hornecker, David Hunger, Jean-Philippe Poizat, Jean-Michel G{\'e}rard, Pascale Senellart, and Alexia Auff{\`e}ves.
\newblock Cavity-funneled generation of indistinguishable single photons from strongly dissipative quantum emitters.
\newblock {\em Physical review letters}, 114(19):193601, 2015.

\bibitem{islam2023cavity}
Fariba Islam, Chang-Min Lee, Samuel Harper, Mohammad~Habibur Rahaman, Yuqi Zhao, Neelesh~Kumar Vij, and Edo Waks.
\newblock Cavity-enhanced emission from a silicon t center.
\newblock {\em Nano Letters}, 24(1):319--325, 2023.

\bibitem{higginbottom2023integrated}
DB~Higginbottom, A~DeAbreu, C~Bowness, A~Alizadeh, C~Chartrand, NA~Brunelle, ER~MacQuarrie, NR~Lee-Hone, M~Ruether, M~Kazemi, et~al.
\newblock Integrated silicon t centers for quantum technologies.
\newblock In {\em Quantum Computing, Communication, and Simulation III}, volume 12446, pages 165--173. SPIE, 2023.

\bibitem{lee2023high}
Chang-Min Lee, Fariba Islam, Samuel Harper, Mustafa~Atabey Buyukkaya, Daniel Higginbottom, Stephanie Simmons, and Edo Waks.
\newblock High-efficiency single photon emission from a silicon t-center in a nanobeam.
\newblock {\em ACS Photonics}, 10(11):3844--3849, 2023.

\bibitem{chang2021detecting}
J~Chang, JWN Los, JO~Tenorio-Pearl, Niels Noordzij, R~Gourgues, A~Guardiani, JR~Zichi, SF~Pereira, HP~Urbach, Val Zwiller, et~al.
\newblock Detecting telecom single photons with 99.5- 2.07+ 0.5\% system detection efficiency and high time resolution.
\newblock {\em APL Photonics}, 6(3), 2021.

\bibitem{reddy2020superconducting}
Dileep~V Reddy, Robert~R Nerem, Sae~Woo Nam, Richard~P Mirin, and Varun~B Verma.
\newblock Superconducting nanowire single-photon detectors with 98\% system detection efficiency at 1550 nm.
\newblock {\em Optica}, 7(12):1649--1653, 2020.

\bibitem{hu2020detecting}
Peng Hu, Hao Li, Lixing You, Heqing Wang, You Xiao, Jia Huang, Xiaoyan Yang, Weijun Zhang, Zhen Wang, and Xiaoming Xie.
\newblock Detecting single infrared photons toward optimal system detection efficiency.
\newblock {\em Optics Express}, 28(24):36884--36891, 2020.

\bibitem{jenkins2022ytterbium}
Alec Jenkins, Joanna~W Lis, Aruku Senoo, William~F McGrew, and Adam~M Kaufman.
\newblock Ytterbium nuclear-spin qubits in an optical tweezer array.
\newblock {\em Physical Review X}, 12(2):021027, 2022.

\bibitem{bowness2025laser}
Camille Bowness, Simon~A Meynell, Michael Dobinson, Chloe Clear, Kais Jooya, Nicholas Brunelle, Mehdi Keshavarz, Katarina Boos, Melanie Gascoine, Shahrzad Taherizadegan, et~al.
\newblock Laser-induced spectral diffusion and excited-state mixing of silicon t centres.
\newblock {\em arXiv preprint arXiv:2504.09908}, 2025.

\bibitem{ngan2024performance}
Kinfung Ngan and Shuo Sun.
\newblock Performance analysis of different photon-mediated entanglement generation schemes under optical dephasing and spectral diffusion.
\newblock {\em arXiv preprint arXiv:2412.09976}, 2024.

\bibitem{Faezehcavityerratum}
F~Kimiaee Asadi, SC~Wein, and Christoph Simon.
\newblock Erratum to cavity-assisted controlled phase-flip gates.
\newblock erratum in preparation, 2025.

\bibitem{duan2004scalable}
L-M Duan and HJ~Kimble.
\newblock Scalable photonic quantum computation through cavity-assisted interactions.
\newblock {\em Physical review letters}, 92(12):127902, 2004.

\bibitem{duan2005robust}
L-M Duan, B~Wang, and HJ~Kimble.
\newblock Robust quantum gates on neutral atoms with cavity-assisted photon scattering.
\newblock {\em Physical Review A}, 72(3):032333, 2005.

\bibitem{xiao2004realizing}
Yun-Feng Xiao, Xiu-Min Lin, Jie Gao, Yong Yang, Zheng-Fu Han, and Guang-Can Guo.
\newblock Realizing quantum controlled phase flip through cavity qed.
\newblock {\em Physical Review A}, 70(4):042314, 2004.

\bibitem{PhysRevA.94.043807}
Chris O'Brien, Tian Zhong, Andrei Faraon, and Christoph Simon.
\newblock Nondestructive photon detection using a single rare-earth ion coupled to a photonic cavity.
\newblock {\em Phys. Rev. A}, 94:043807, Oct 2016.

\bibitem{walls2008quantum}
DF~Walls and Gerard~J Milburn.
\newblock Quantum information.
\newblock In {\em Quantum Optics}, pages 307--346. Springer, 2008.

\bibitem{vandersypen2004nmr}
Lieven~MK Vandersypen and Isaac~L Chuang.
\newblock Nmr techniques for quantum control and computation.
\newblock {\em Reviews of modern physics}, 76(4):1037--1069, 2004.

\bibitem{hill2015surface}
Charles~D Hill, Eldad Peretz, Samuel~J Hile, Matthew~G House, Martin Fuechsle, Sven Rogge, Michelle~Y Simmons, and Lloyd~CL Hollenberg.
\newblock A surface code quantum computer in silicon.
\newblock {\em Science advances}, 1(9):e1500707, 2015.

\bibitem{clear2024optical}
Chloe Clear, Sara Hosseini, Amirhossein AlizadehKhaledi, Nicholas Brunelle, Austin Woolverton, Joshua Kanaganayagam, Moein Kazemi, Camille Chartrand, Mehdi Keshavarz, Yihuang Xiong, et~al.
\newblock Optical-transition parameters of the silicon t center.
\newblock {\em Physical Review Applied}, 22(6):064014, 2024.

\bibitem{bergeron2019optical}
Laurent Bergeron.
\newblock Optical and magnetic properties of the t radiation damage centre in $^{28}$si.
\newblock Master's thesis, Simon Fraser University, 2019.

\bibitem{janitz2020cavity}
Erika Janitz, Mihir~K Bhaskar, and Lilian Childress.
\newblock Cavity quantum electrodynamics with color centers in diamond.
\newblock {\em Optica}, 7(10):1232--1252, 2020.

\bibitem{o2016nondestructive}
Chris O'Brien, Tian Zhong, Andrei Faraon, and Christoph Simon.
\newblock Nondestructive photon detection using a single rare-earth ion coupled to a photonic cavity.
\newblock {\em Physical Review A}, 94(4):043807, 2016.

\bibitem{thiele2025cryogenic}
Frederik Thiele, Niklas Lamberty, Thomas Hummel, Nina~A Lange, Lorenzo~M Procopio, Aishi Barua, Sebastian Lengeling, Viktor Quiring, Christof Eigner, Christine Silberhorn, et~al.
\newblock Cryogenic feedforward of a photonic quantum state.
\newblock {\em Optica}, 12(5):720--727, 2025.

\bibitem{metz2008effect}
J~Metz and SD~Barrett.
\newblock Effect of frequency-mismatched photons in quantum-information processing.
\newblock {\em Physical Review A—Atomic, Molecular, and Optical Physics}, 77(4):042323, 2008.

\bibitem{kambs2018limitations}
Benjamin Kambs and Christoph Becher.
\newblock Limitations on the indistinguishability of photons from remote solid state sources.
\newblock {\em New Journal of Physics}, 20(11):115003, 2018.

\bibitem{uysal2024rephasing}
Mehmet~T Uysal and Jeff~D Thompson.
\newblock Rephasing spectral diffusion in time-bin spin-spin entanglement protocols.
\newblock {\em arXiv preprint arXiv:2406.06497}, 2024.

\bibitem{zhang2025laser}
Xueyue Zhang, Niccolo Fiaschi, Lukasz Komza, Hanbin Song, Thomas Schenkel, and Alp Sipahigil.
\newblock Laser-induced spectral diffusion of t centers in silicon nanophotonic devices.
\newblock {\em arXiv preprint arXiv:2504.08898}, 2025.

\bibitem{lago2021telecom}
Dario Lago-Rivera, Samuele Grandi, Jelena~V Rakonjac, Alessandro Seri, and Hugues de~Riedmatten.
\newblock Telecom-heralded entanglement between multimode solid-state quantum memories.
\newblock {\em Nature}, 594(7861):37--40, 2021.

\bibitem{mcauslan2009strong}
DL~McAuslan, Jevon~Joseph Longdell, and MJ~Sellars.
\newblock Strong-coupling cavity qed using rare-earth-metal-ion dopants in monolithic resonators: What you can do with a weak oscillator.
\newblock {\em Physical Review A—Atomic, Molecular, and Optical Physics}, 80(6):062307, 2009.

\bibitem{asadi2018quantum}
F~Kimiaee Asadi, N~Lauk, S~Wein, N~Sinclair, C~O'Brien, and C~Simon.
\newblock Quantum repeaters with individual rare-earth ions at telecommunication wavelengths.
\newblock {\em Quantum}, 2:93, 2018.

\bibitem{ji2022proposal}
Jia-Wei Ji, Yu-Feng Wu, Stephen~C Wein, Faezeh~Kimiaee Asadi, Roohollah Ghobadi, and Christoph Simon.
\newblock Proposal for room-temperature quantum repeaters with nitrogen-vacancy centers and optomechanics.
\newblock {\em Quantum}, 6:669, 2022.

\bibitem{rozpkedek2019near}
Filip Rozp{\k{e}}dek, Raja Yehia, Kenneth Goodenough, Maximilian Ruf, Peter~C Humphreys, Ronald Hanson, Stephanie Wehner, and David Elkouss.
\newblock Near-term quantum-repeater experiments with nitrogen-vacancy centers: Overcoming the limitations of direct transmission.
\newblock {\em Physical Review A}, 99(5):052330, 2019.

\bibitem{main2025distributed}
D~Main, P~Drmota, DP~Nadlinger, EM~Ainley, A~Agrawal, BC~Nichol, R~Srinivas, G~Araneda, and DM~Lucas.
\newblock Distributed quantum computing across an optical network link.
\newblock {\em Nature}, pages 1--6, 2025.

\bibitem{barral2024review}
David Barral, F~Javier Cardama, Guillermo D{\'\i}az, Daniel Fa{\'\i}lde, Iago~F Llovo, Mariamo~Mussa Juane, Jorge V{\'a}zquez-P{\'e}rez, Juan Villasuso, C{\'e}sar Pi{\~n}eiro, Natalia Costas, et~al.
\newblock Review of distributed quantum computing. from single qpu to high performance quantum computing.
\newblock {\em arXiv preprint arXiv:2404.01265}, 2024.

\end{thebibliography}

\newpage

\appendix
\renewcommand{\appendixname}{}  
\renewcommand{\thesection}{S\Roman{section}}  
\renewcommand{\theequation}{S\arabic{equation}}  
\setcounter{equation}{0}  
\renewcommand{\thefigure}{\thesection.\arabic{figure}}
\setcounter{figure}{0} 

\section*{Supplementary Material}

\section{INTERFERENCE-BASED SCHEME WITH FEEDBACK, INCLUDING THE EFFECT OF OPTICAL FREQUENCY MISMATCH AND PHASE ERRORS}
\label{IBF no spin dephasing}
If the gate time is significantly shorter than the spin coherence time, the effect of spin decoherence can be neglected and the fidelity of the scheme can be calculated analytically considering the optical frequency difference between the two T centers, and phase errors. The fidelity of the IBF scheme is given by

\begin{equation}
F_{\text{IBF}}=\frac{1}{2}\left(1+\frac{\Re(\Tilde{C}_{\text{IBF}}(T_d))}{1-e^{-\gamma'T_d}}\right),
\label{FIBF}
\end{equation}
where $\Re(\Tilde{C}_{\text{IBF}}(T_d))$ is the real part of the $\Tilde{C}_{\text{IBF}}(T_d)$ that is
\begin{equation}
\Tilde{C}_{\text{IBF}}(T_d)=\frac{\gamma'}{\Gamma'+i\Delta}e^{-\delta t(2\gamma^*+i\Delta)}(1-e^{-T_d(\Gamma'+i\Delta)})e^{i(\varphi+\phi)}.    
\end{equation} 
Here $\Gamma' = \gamma' + 2\gamma^*$ represents the FWHM of the Purcell-enhanced emission line, $\gamma^*$ is the optical pure dephasing rate which is $\gamma^* = 1/T_{2h} - \gamma/2$, $\Delta$ is the optical frequency difference between the two T centers, $\varphi = \varphi_1 - \varphi_2$ is the relative initialization phase, and $\phi = \phi_1 - \phi_2$ is the relative propagation phase \cite{wein2020analyzing}. Infidelity arises from optical decay and dephasing rates, frequency mismatch between the two T centers, and phase errors, all of which affect photon indistinguishability. In the presence of spin dephasing, and assuming there is no spectral detuning $\Delta=0$ and relative initial phases $\varphi=\phi=0$ between the T centers, the fidelity of the scheme can be calculated analytically as shown in the main text. 

\section{ELECTRIC DIPOLE GATE}
\label{elecgate}
\begin{figure}[h!]
\centering
\includegraphics[scale=0.3]{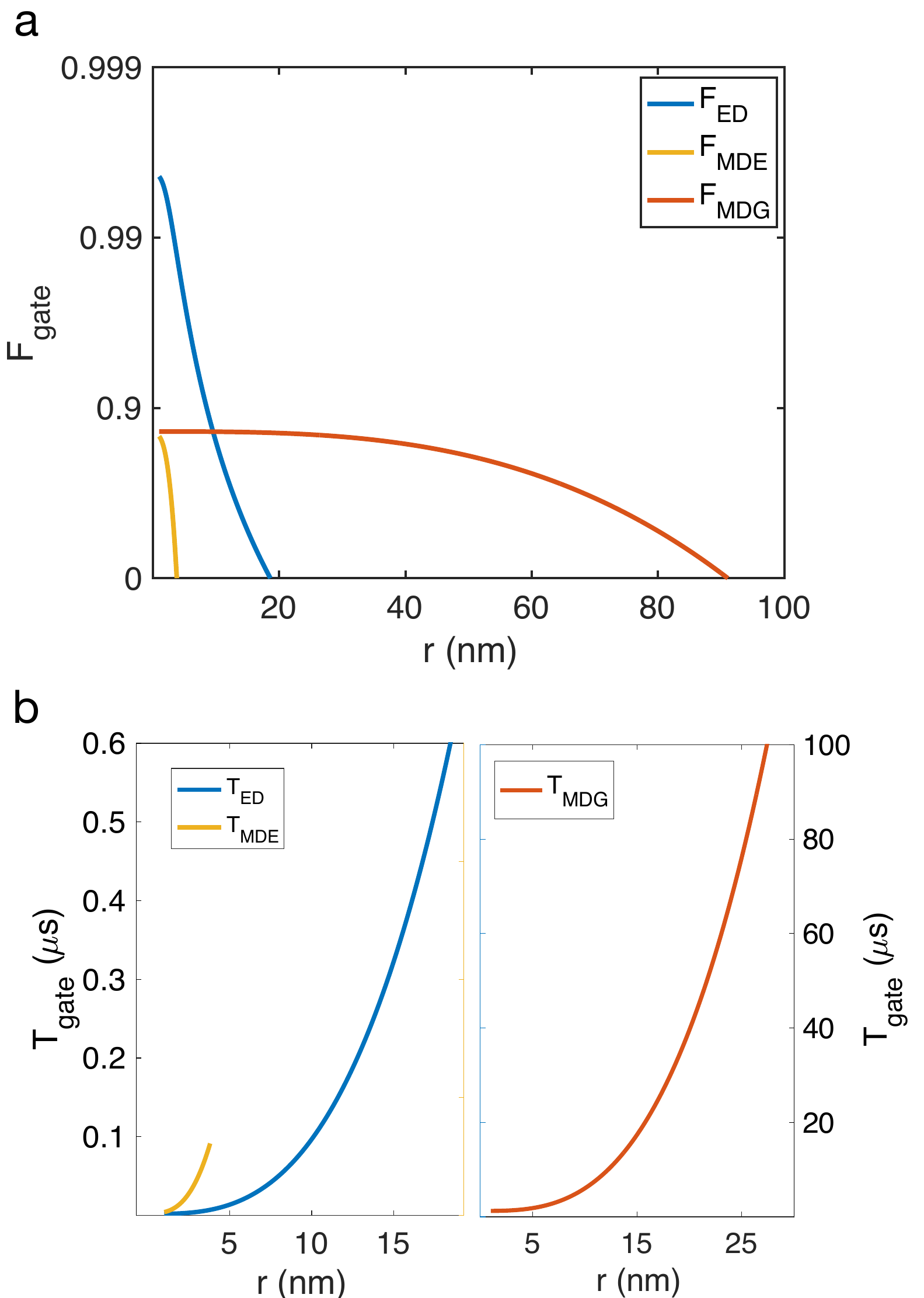}
\caption{Comparison of the \textbf{(a)} fidelity $(F_{gate})$ and \textbf{(b)} gate time $(T_{gate})$ for the electric dipole gate (ED), and the magnetic dipole gate (performing in both the excited (MDE) and ground state (MDG)) with respect to the distance $r$ between the two T centers. For each gate, only the variable ranges that yield meaningful values for fidelity are shown.}
\label{FTdistancedep}
\end{figure}

Electric dipole (ED) based gates have been investigated in rare-earth-ion-doped crystals \cite{ohlsson2002quantum}, and here we adapt this approach for the T center. The lack of site symmetry leads the color centers belonging to some crystallographic point groups to have a permanent electric dipole moment, which changes when a T center is excited from the ground state to the $\mathrm{{TX}_0}$ excited state. The transition frequency of the nearby individual T center is affected by the modification in the environmental electric field which is the result of changing in the permanent electric dipole moment of the neighbouring T center excitation. Therefore, by optical excitation of the T center one can control the shift in the transition frequency of the neighbouring center and perform a controlled-NOT (CNOT) operation between nearby T centers. The required pulse sequences and how to perform the gate are detailed in \cite{ohlsson2002quantum}. The change in the transition frequency $(\Delta\nu)$ of the nearby T center is estimated by \cite{ohlsson2002quantum, asadi2018quantum}
\begin{equation}
\Delta \nu = \frac{\Delta \mu_{T} \Delta \mu_{T}}{4 \pi \epsilon \epsilon_0 h r^3} \left( (\hat{\mu}_{T} \cdot \hat{\mu}_{T}) - 3 (\hat{\mu}_{T} \cdot \hat{r}) (\hat{\mu}_{T} \cdot \hat{r}) \right)
\label{Deltanu}
\end{equation}
where $\Delta\mu_{T}$ is the
change in the permanent electric dipole moment of the T center with the measured value of $\Delta\mu_{T} = 49\times10^{-31}\,\mathrm{Cm}$ \cite{clear2024optical}, $r$ is the separation between the two T centers, $h$ and $\epsilon_0$ are the Planck constant and the vacuum permittivity, and $\epsilon$ is the dielectric constant. With the refractive index of $n=3.45$ for silicon, one can calculate the dielectric constant $\epsilon = n^2$. Therefore, $\Delta\nu$ can be obtained for different T centers' separations. The shift in the transition frequency $(\Delta\nu)$ should be more than a few homogeneous linewidth which is $\Gamma^{hom}/2\pi = 690 \,\mathrm{kHz}$ for the T center \cite{deabreu2023waveguide}. One should note that here we consider only the $y$-component for the calculation of $\Delta\mu_{T}$ in equation (6) of \cite{clear2024optical}. However, if both components are considered, we obtain $\Delta\mu_{T}=55\times10^{-31}\,\mathrm{Cm}$, corresponding to larger distances between the T centers for similar values of $\Delta\nu$ (see equation \eqref{Deltanu}.    
\\\indent To perform the gate we need three levels (see figure $2$ in \cite{asadi2018quantum}). For T center we can work with the two electron spin levels $\ket{\uparrow_{e}}$ and $\ket{\downarrow_{e}}$, and one of the hole spin levels $\ket{\downarrow_{h}}$ in figure \ref{Tcenter}\,\textbf{b}. In total, we apply five pulses with a Rabi frequency of $\Omega$ to perform the gate: two pulses on the T center acting as the control qubit ($\Omega_c$) and three pulses on the T center acting as the target qubit ($\Omega_t$) (see figure 2 in \cite{asadi2018quantum}). Consequently, the total gate time for the electric dipole gate is given by $T_{\mathit{ED}}= \frac{2\pi}{\Omega_c}+\frac{3\pi}{\Omega_t}$.  
\\\indent For a high-fidelity gate, it is essential to ensure that the frequency shift $\Delta\nu$ is sufficiently large to prevent the lasers from exciting the frequency-shifted T center (the target qubit) while the neighboring T center (the control qubit) is being excited. This requires the condition ${\Omega_t}^2 \ll \Delta\nu^2$. This leads to a very slow gate and a low fidelity because $\Delta\nu$ cannot be arbitrarily large (the distance between T centers cannot be arbitrarily small). To overcome this issue, one can perform an effective $2\pi$ pulse on the T center spin state when the nearby T center is excited as suggested in rare-earth-ion-doped crystals \cite{asadi2018quantum}. Then $\Omega_t = \Delta\nu/\sqrt{3}$ and $T_{\mathit{ED}} = \frac{2\pi}{\Omega_c}+\frac{3\pi\sqrt{3}}{\Delta\nu}$. Also, to ensure that only one of the electron spin levels is addressed by the applied pulses, $\Omega$ should be smaller than the frequency splitting between the electron spin levels. For example for a splitting of $\Delta_e = 2\pi \times 2.25\, \mathrm{GHz}$, a ratio of $\frac{\Delta_e}{\Omega}=4$ corresponds to an error of $10^{-11}$ in the fidelity due to unwanted off-resonant excitations (see section \ref{MDG}). The fidelity of the electric dipole gate, determined by analytically solving the master equation for a nine-level system (with each T center modeled as a three-level system), is given by \cite{asadi2018quantum}

\begin{equation}
F_{\text{ED}} = 1 - \frac{\mathrm{T_{\text{ED}}}}{80}(42\gamma + 25 \gamma^* + 25 \chi) - \frac{43\pi^2}{128}(\frac{\delta\nu}{\Delta\nu})^2
\label{FED}
\end{equation}
where $\gamma = 1/T_{1h}$ is the optical decay rate, $\gamma^*$ is the optical pure dephasing rate which is $\gamma^* = 1/T_{2h} - \gamma/2$ for the optical transition, $\chi = 1/T_{2e}$ is the electron spin decoherence rate, and $\delta\nu$ is a small error from the true value of $\Delta\nu$.
\\\indent Figure \ref{FTdistancedep}\,\textbf{a} (\textbf{b}) shows the fidelity $F_{\mathit{ED}}$ (gate time $T_{\mathit{ED}}$) with respect to the distance $r$ between the T centers using equations \eqref{Deltanu} and \eqref{FED}. Note $T_{\mathit{ED}}$ should be less than the optical lifetime and dephasing time. For reasonable gate times and improved fidelity, $\Delta \nu$ must be increased, i.e., the separation between T centers must be decreased, which is currently beyond the reach of existing technology. However, if T centers could be created close enough to achieve a sufficiently large $\Delta \nu$, performing this gate would become feasible.

\section{MAGNETIC DIPOLE GATE in the EXCITED STATES}
\label{MDE}
Here, we consider performing the magnetic dipole gate in the excited states $(MDE)$, selecting the electron spin states in the ground state as the "passive" qubits and the hole spin states in the excited state as the "active" qubits. The anisotropy of the unpaired hole spin in the bound exciton excited state results in 12 independently addressable orientational subsets, meaning each T center would have a specific orientation \cite{bergeron2020silicon} (see figure \ref{Tcenter}\,\textbf{b}). We choose the first orientational subset as it has the largest g-value. To calculate the gate time and fidelity, one needs the information about the g-tensor in the excited state, but such information is not available. Assuming $g_x = g_y = g_z = 3.45$ \cite{bergeron2020silicon} and a decay time of $100 \, \mathrm{ns}$ between the hole spins up and down, figure \ref{FTdistancedep}\,\textbf{a} (\textbf{b}) shows the fidelity $F_{\mathit{MDE}}$ (gate time $T_{\mathit{MDE}}$) of the magnetic dipole gate with respect to the separation between the T centers (see section \ref{MDG} for equations). Note the gate time $T_{\mathit{MDE}}$ should be shorter than the optical decay time, the decoherence time between the excited states (hole spins), and the decoherence time between the ground states. The assumption of $g_x = g_y = g_z$ leads to a second-order error contribution of $0.02$ to the fidelity. The infidelity term mentioned in section \ref{MDG}, which arises from unwanted off-resonant excitations during the application of $\pi$ pulses with Rabi frequency $\Omega$, is approximately $10^{-13}$. It’s important to keep in mind that this result excludes terms involving the excited-state spin dephasing rate $\gamma_5$ in the fidelity equation \eqref{FMD}, as there are no measurements of this parameter for the T center. However, performing this gate in the excited states is generally unattractive, as achieving high fidelity and reasonable gate times would require the distance between the two T centers to be less than $2\, \mathrm{nm}$. Therefore, implementing this gate in the excited state (hole spins) is not feasible with current technology. 

\section{SIMPLE VIRTUAL PHOTON EXCHANGE}
\label{simple}
\begin{figure}[h!]
\centering
\includegraphics[scale=0.42]{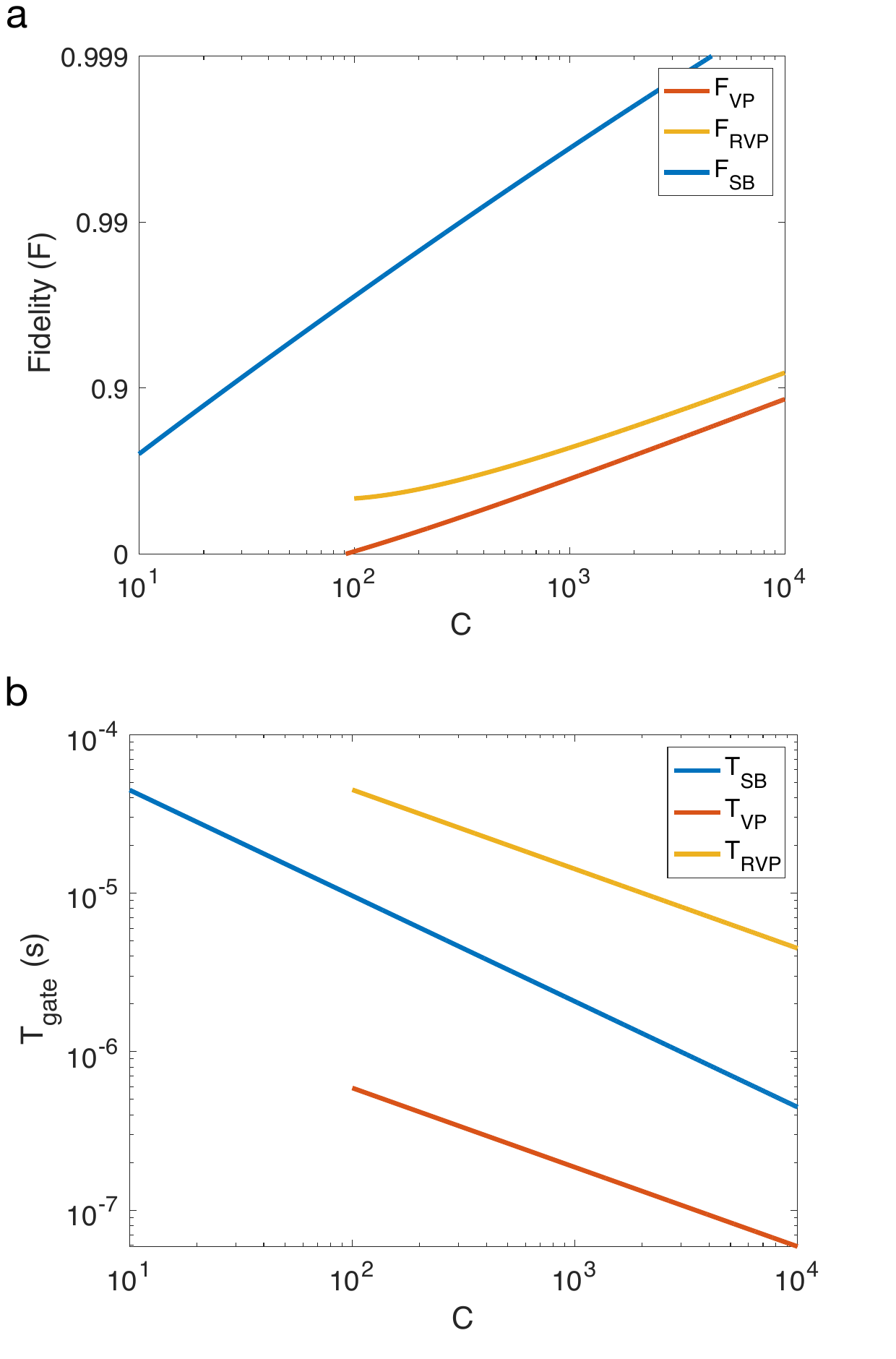}
\caption{Comparison of the \textbf{(a)} fidelity $(F_{gate})$ and \textbf{(b)} gate time $(T_{gate})$ for the photon scattering-based (SB), simple virtual photon exchange (VP), and Raman virtual photon exchange (RVP) schemes with respect to the cavity cooperativity $C$ for implementation in the T center. For the VP scheme, it is assumed that ${\Delta_\epsilon = 0}$, with the relevant splittings in the ground and excited states for the T center given as $\delta_{eg} = 1.62\,\mathrm{GHz}$. For the RVP scheme, ${\Delta_\epsilon = \delta_\epsilon = 0}$ and $\Omega = 0.1 \Delta$ are assumed. For the SB scheme, we assume ${\delta_p = \delta_{\epsilon_{A}} = \delta_{\epsilon_{B}} = 0}$, with $g/\kappa = 0.01$ based on experimental data \cite{johnston2024cavity}. For each gate, only the variable ranges that yield meaningful values for fidelity are shown.}
\label{FTCdep}
\end{figure}

The simple virtual photon exchange scheme (VP) is based on a cavity-assisted interaction between qubits. A phase-flip gate is performed when the two systems’ optical transitions are in resonance but dispersively coupled to a cavity mode. 
\\\indent Using the non-Hermitian Hamiltonian approach and solving the effective non-Hermitian component of the master equation yields an analytical expression for the fidelity of the simple virtual photon exchange gate that is well approximated by \cite{asadi2020cavity, asadi2020protocols}

\begin{multline}
F_{\text{VP}} = 1 - \frac{2\pi}{\sqrt{C}} - \Gamma \mathrm{T_{VP}} - 0.58 \mathrm{T_{VP}}\gamma^* \\ - \frac{6\pi^2}{32}\left[\left(\frac{\mathrm{T_{VP}}\Delta_\epsilon}{2\pi}\right)^2 + \left(\frac{2\pi}{\mathrm{T_{VP}}\delta_{eg}}\right)^2 - \frac{12}{C}\right] 
\label{fidelity simple}    
\end{multline}

where $\Delta_\epsilon$ is the detuning between the quantum systems’ optical transitions which is small, $\delta_{eg}$ is the difference between ground-state and excited-state splittings, $\Gamma = 1/T_{2e}$ is the effective decoherence rate for the ground state, $\gamma^* = 1/T_{2h} - \gamma/2$ is the optical pure dephasing rate, and $T_{\mathit{VP}}$ is the gate time of VP scheme. Here, we consider the optimal gate time using the equation $T_{\mathit{VP}}=2\pi/\gamma\sqrt{C}$, which is achieved under the condition that maximizes the gate fidelity \cite{asadi2020cavity}. In this equation, $\gamma= 1/T_{1h}$ is the optical decay rate, and $C$ is the cavity cooperativity. Note that equation \eqref{fidelity simple} includes a factor of $2$ in the infidelity terms compared to the fidelity equation in \cite{asadi2020cavity}, as the fidelity definition used here is the square of the definition used in Ref. \cite{asadi2020cavity}. To account for the effect of optical pure dephasing, the analytical approximation is compared with the numerical solution from simulating the master equation in the bad-cavity regime and the coefficient of $0.58$ in the third infidelity term is obtained \cite{asadi2020protocols}. To achieve high fidelity the gate time $T_{\mathit{VP}}$ must be shorter than the excited state lifetime $T_{1h}$, and the system should avoid entering the strong-coupling regime $g/\kappa \leq 1$ \cite{asadi2020cavity}. 
\\\indent Four levels are required to perform the gate. We employ the two electron spin states in the ground state $\ket{\downarrow_{e}}$, $\ket{\uparrow_{e}}$ and the two hole spin states in the bound exciton excited state $\ket{\downarrow_{h}}$, $\ket{\uparrow_{h}}$ to perform the gate. Figure \ref{FTCdep}\,\textbf{a} (\textbf{b}) shows fidelity $F_{\mathit{VP}}$ (gate time $T_{\mathit{VP}}$) for VP scheme. The result shows obtaining favorable gate time and fidelity for the VP gate in T centers requires relatively large cavity cooperativity. Considering the current experimental value for the cavity cooperativity of the T center \cite{johnston2024cavity}, the largest realistic value for $C$ in the foreseeable future would be around $100$. Therefore, this gate would not be suitable for the T center in the near term. With the advent of cavity fabrication technology and increase in the $C$ value this gate might be practicable for implementation on T center.

\section{RAMAN VIRTUAL PHOTON EXCHANGE}
\label{Raman}
Raman virtual photon exchange scheme (RVP) is another phase-flip gate between distant qubits that can be performed using virtual excitation of the cavity mode via a Raman coupling. This gate has been implemented on quantum dots and trapped ions. The proposed scheme in \cite{asadi2020cavity} helps to overcome the challenges of the previous proposals and also allows one to perform a phase-flip gate between
qubits in systems that have unequal optical transitions. 
\\\indent We consider two 4-level systems
where the systems are dispersively coupled to a
far-detuned cavity. The levels to perform the gate in T centers are the nuclear spins associated with the electron spins in the ground state $\ket{1}$, $\ket{3}$, and $\ket{4}$ for shelving, and hole spin down $\ket{\downarrow_{h}}$ in the excited state. 
\\\indent The RVP scheme is analyzed similarly to the VP scheme, with the fidelity calculated by solving the master equation using the non-Hermitian Hamiltonian method and is well approximated by \cite{asadi2020cavity}

\begin{multline}
F_{\text{RVP}} = 1 - \frac{2\pi}{\sqrt{C}} - \Gamma \mathrm{T_{RVP}} \\ - \frac{\pi^2}{8}\left[\left(\frac{\mathrm{T_{RVP}}\delta_\epsilon}{2\pi}\right)^2 + \frac{\Delta_\epsilon^2}{\Delta^2} - \frac{18}{C}\right] 
\label{fidelity raman}
\end{multline}

where $\delta_\epsilon$ is a small two-photon resonance error, $\Delta$ is the detuning between the cavity mode and the resonance transition of the quantum system, $\Gamma = 1/T_{2e}$ is the effective decoherence rate for the ground state, $\Delta_\epsilon$ is a small detuning between the optical frequencies of the systems, and $T_{\mathit{RVP}}$ is the gate time of RVP scheme. Under the condition that maximizes the scheme fidelity \cite{asadi2020cavity}, the optimal gate time is given by $T_{\mathit{RVP}}=(\Delta/\Omega)^2\,(2\pi/\gamma\sqrt{C})$, where $\Omega$ is the Rabi frequency of the classical laser field. To achieve high fidelity the gate time $T_{\mathit{RVP}}$ should be shorter than the lifetime of the shelved state, the system should remain in the bad-cavity regime $g < \kappa$, and $\Omega \ll \Delta$ to avoid populating the excited state. Although in the RVP scheme the excited state is not populated, the fidelity in equation \eqref{fidelity raman} represents an upper bound. 
\\\indent To approximately account for the detrimental effect of optical pure dephasing on fidelity, we adopt the approach in Ref. \cite{karimi2024comparing}, which quantifies this effect by replacing the cavity cooperativity $C$ with an effective cooperativity $C_{\mathit{eff}} = C / [1 + 0.7\,(T_{1h} / T_{2h} - 1)]$ in equation \eqref{fidelity raman}, indicating that optical pure dephasing effectively reduces the cavity cooperativity. Note that $T_{\mathit{RVP}}$ is also dependent on the cavity cooperativity $C$, so one should consider the change in $T_{\mathit{RVP}}$ due to optical pure dephasing when calculating the fidelity \cite{asadi2020protocols}.
\\\indent Figure \ref{FTCdep}\,\textbf{a} (\textbf{b}) shows $F_{\mathit{RVP}}$ ($T_{\mathit{RVP}}$) for RVP scheme. Employing the same values of cavity cooperativity $C$ as the VP scheme one can achieve higher fidelity by performing the Raman scheme compared to the simple virtual scheme. However, even for the high cooperativities the time for performing the Raman gate $T_{\mathit{RVP}}$ is much slower. Although a slow gate is a challenge if one wants to integrate the system in a chain of other operations, for Raman virtual gate there is no real excitation of the atoms, and so there is no limitation on the gate time compared to the simple virtual gate where the gate time should be less than the optical decoherence time.

\end{document}